\documentclass[aps, preprint]{revtex4}
\usepackage{graphicx}
\usepackage{epsf}
\usepackage{wrapfig}
\usepackage{epsfig}
\def\bk{{\mbox{\boldmath$k$}}}
\def\ner #1{\mbox{\boldmath$ #1$}}
\def\bq{{\mbox{\boldmath$q$}}}

\def\bp{{\mbox{\boldmath$p$}}}

 \def\br{{\mbox{\boldmath$r$}}}

   \def\bp{{\mbox{\boldmath$p$}}}

\def\br{{\mbox{\boldmath$r$}}}
\def\b0{{\mbox{\boldmath$0$}}}
\def\dfrac{\displaystyle\frac}
\def\bk{{\mbox{\boldmath$k$}}}
\def\bq{{\mbox{\boldmath$q$}}}

\def\bp{{\mbox{\boldmath$p$}}}

\def\br{{\mbox{\boldmath$r$}}}
\def\b0{{\mbox{\boldmath$0$}}}

\newcommand{\ra}{\,\rangle}
\newcommand{\la}{\,\langle}

 \newcommand{\CE}{{E}}

\def \b #1{ {\bf #1}}
\newcommand{\be}{\begin{eqnarray}}
\newcommand{\ee}{\end{eqnarray}}

\def\dfrac{\displaystyle\frac}

     \font\tenbifull=cmmib10 scaled 1200 
     \font\tenbimed=cmmib9
     \font\tenbismall=cmmib7
       \def\bmit{\fam9 }
\mathchardef\bbkappa="7114
\mathchardef\bbrho="711A
\mathchardef\bbsigma="711B
\mathchardef\bbtau="711C
\mathchardef\bbvarrho="7125
\mathchardef\bbvarsigma="7126
\mathchardef\bbxi="7118
\def\boldkappa{{\bmit\bbkappa}}

\def\boldtau{{\bmit\bbtau}}

\usepackage{graphicx}
\usepackage{epsf}
\usepackage{wrapfig}
\usepackage{epsfig}
\def\Vec#1{\mbox{\boldmath $#1$}}

\def\dfrac{\displaystyle\frac}

\def\n{\noindent}
\def\beq{\begin{equation}}
\def\eeq{\end{equation}}

\def\beqy{\begin{eqnarray}}
\def\eeqy{\end{eqnarray}}

\newcommand{\ber}{\begin{displaymath}}
\newcommand{\eer}{\end{displaymath}}

\begin{document}
\hyphenation{author another created financial paper re-commend-ed
Post-Script}
\title{
Final State Interaction Effects in Semi-inclusive Deep Inelastic processes $A(e,e'p)X$ off
the deuteron and
complex nuclei}

 \author{M. Alvioli}\author{C. Ciofi degli Atti}\author{V. Palli}
  \author{L.P. Kaptari}
  \altaffiliation{On leave from  Bogoliubov Lab.
      Theor. Phys., 141980, JINR,  Dubna, Russia}
      \address{Department of Physics, University of Perugia and
      Istituto Nazionale di Fisica Nucleare, Sezione di Perugia,
      Via A. Pascoli, I-06123, Italy\\
      (\today)}

\begin{abstract}
\n
  The effects of the final state interaction (FSI) in semi inclusive deep inelastic
  electron scattering  processes $A(e,e'p)X$  off nuclei  are investigated in details.
  Proton production is described  within the
  spectator and the target fragmentation mechanisms whose
  relevance to the experimental
  study of the deep inelastic structure functions of bound nucleons and  the
  non perturbative hadronization process is analyzed.
  Particular attention is paid to the deuteron target within
  kinematical conditions corresponding to the available and forthcoming
   experimental data  at Jlab.
 We argue that there are kinematical regions where  FSI effects are
 minimized,  allowing for a reliable investigation of the DIS structure functions,
  and regions where the interaction of the quark-gluon debris with nucleons  is maximized,
  which makes it possible to study  hadronization mechanisms.
Nuclear structure has been described by means of realistic
   wave functions and spectral functions and the  final state interaction   has been
  treated within    an eikonal approximation approach which takes into account
   the rescattering of the quark-gluon debris with  the residual
  nucleus  and,    in
  the case of complex nuclei, within an optical potential approach
    to account for  the FSI  of the struck proton.
\end{abstract}
\maketitle

\section{Introduction}
\n Semi inclusive Deep Inelastic Scattering (SIDIS) of leptons
off nuclei
 can provide relevant
information on: (i) possible modification of the nucleon structure function in
 medium (EMC-like effects),
  (ii) the relevance of exotic configurations at short NN distances;
  (iii) the mechanism of quark hadronization
  (see e.g. \cite{FS}).
   A process which attracted much interest from both the theoretical \cite{review,ciosim2}
  and experimental \cite{experiment}
  points of view is the production of slow
  protons,  i.e. the process  A({\cal l},{\cal l}'p)X, where a slow proton is detected
  in coincidence
with the scattered lepton.
In plane wave impulse approximation (PWIA),
  after the hard collision of $\gamma^\star$ with
  a quark of a bound nucleon, two main mechanisms of production of slow protons
  have been considered, namely the spectator mechanism \cite{FS},
   and target fragmentation \cite{review,ciosim2}.
   In the target fragmentation slow protons originate
   from  the spectator diquark which captures a quark from the vacuum
   creating the recoiling proton. In case of fragmentation of
   free nucleons by  highly virtual photons this is the only source of
   slow nucleons.
   In SIDIS off nuclei both  target fragmentation and the spectator processes can occur,
   the latter  originating from   nucleon-nucleon (NN) short range correlations.
    In the past the spectator mechanism has been intensively investigated,
    though  most calculations either completely disregarded  the effects of the final state
    interaction,  or considered them within simple models. In this paper
    the  results of calculations of both the spectator and fragmentation mechanisms,
    taking into account FSI effects will be presented. Our paper is organized as follows:
     the results
    for the deuteron are presented in Section II, complex nuclei are treated in Section III,
    the Conclusions
    are presented in Section IV.

\section{Proton production from the deuteron}
\n
 We analyze   the  process of proton production from the deuteron target
of the type
 \beq
 e+D=e'+p+X,
 \label{Dreac}
 \eeq
 which not only has been considered in several theoretical papers \cite{simula,sarg,scopetta},
 but
 is at present under experimental investigation at JLab \cite{bonus}. This process,
 if the spectator mechanism is valid, can provide unique information on the
 DIS structure functions of bound nucleons.
 In ref.~\cite{ckk} the theoretical treatment of the spectator mechanism approach has been
 implemented  by considering the effects of the FSI
 between the hadrons  (mostly mesons), created during the hadronization of
 the quark-gluon debris, and  the spectator proton. Note that
 hadronization  is basically a QCD  nonperturbative  process, and, consequently,
 any   experimental information on its  effects on the reaction (\ref{Dreac})
  would be a rather valuable one.
  It has been observed in ref. \cite{ckk} that in the kinematical range where the FSI effects
  are relevant, the
  process (\ref{Dreac}) is essentially governed by the  hadronization cross section.
 This opens a new and important
  aspect of these reactions, namely  the possibility, through them, to investigate
  hadronization mechanisms
  by choosing a proper kinematics where FSI effects are maximized.
 Herebelow we improve  our previous analysis  of the process
  (\ref{Dreac}) by also considering the possibility
  of proton production due to target fragmentation.
  For the sake of simplicity, we consider here
  DIS kinematics  in the Bjorken limit
  ($Q^2\to \infty; \nu\to\infty$;  $x=\dfrac{Q^2}{2m\nu}$ finite);
  the  generalization to finite values of $Q^2$ is straightforward.

In the one-photon-exchange approximation
the   cross section for the process given by (\ref{Dreac}) can be written as follows
\be &&
\frac{d\sigma}{dx dQ^2\ d\bp_2}=\frac{4\alpha_{em}^2}{Q^4}\frac{\pi\nu}{x}
\left[ 1-y-\frac{Q^2}{4\CE^2}
\right]
\widetilde{l}^{\mu\nu}L_{\mu\nu}^D\equiv
\label{eq1}\\&&
\frac{4\alpha_{em}^2}{Q^4}\frac{\pi\nu}{x}
\left[ 1-y-\frac{Q^2}{4\CE^2}
\right]
\left [
\tilde l_L W_L +\tilde l_{T} W_T+\tilde l_{TL} W_{LT}\cos\phi_s+
\tilde l_{TT} W_{TT}\cos (2\phi_s)
\right ]
\label{cross}
\ee
\noindent where  $\alpha_{em}$ is the fine-structure constant,
 $Q^2 =-q^2= -(k-k')^2 = {\bf q}^{\,\,2} - \nu^2=4\CE\CE'
sin^2 {\theta \over 2}$ the four-momentum transfer (with ${\bf q} =
{\bf k} - {\bf  k}'$, $\nu=   \CE  -  \CE ^{'} $ and $ \theta \equiv
\theta_{\widehat{\vec k \vec k^{'}}}$), $ x = Q^2/2M\nu $ the
Bjorken scaling variable,
 $y=\nu /\CE$,
 ${\bf p}_2$ the momentum of the detected recoiling  nucleon, and $\widetilde {l}_{\mu\nu} $
 and $L_{\mu\nu}^D$ the electron
 and deuteron electromagnetic tensors, respectively;  the former has the well known standard
 form \cite{boffi}, whereas
 the latter can be written as follows
\be
&&
 L_{\mu\nu}^D=
  \sum\limits_{X}\la {\bf P}_D|J_\mu|{\bf P}_f\ra\la {\bf P}_f|J_\nu|{\bf P}_D\ra
(2\pi)^4 \delta^{(4)}\left(k+P_D-k'-p_X-p_2 \right)
 d\boldtau_X,
\label{eq5} \ee where $J_\mu$ is the operator of the deuteron
electromagnetic current, and ${\bf P}_D$ and ${\bf P}_f= {\bf
p}_X+{\bf p}_2$ denote the three-momentum of the initial deuteron
and the final hadron system, respectively, with ${\bf p}_X$ being the
momentum of the undetected hadronic state created by the DIS process on the active nucleon.
 In Eq.
(\ref{cross}) the various $W_i$ represent the nuclear response
functions and the quantities $\tilde l_i$ are the  corresponding  components of the photon
spin density matrix  \cite{boffi}.
It is well known, that within the PWIA, i.e., when FSI effects are disregarded,
the four response functions in Eq. (\ref{cross}) can be expressed in terms of
 two independent structure functions, {\it viz}  $W_L$ and $W_T$.
 Moreover, in the DIS kinematics, when
  the Callan-Gross  relation holds, $2x F_1^N(x) =  F_2^N(x)$,  the semi inclusive
   cross section (\ref{cross}) will depend
   only upon one DIS structure function, e.g.
    $F_2^N(x)$.  In the presence of  FSI all four responses contribute to the cross section
    (\ref{cross});
  however if  FSI effects  are not too large, nucleon
 momenta are sufficiently small ($|\bp_2|^2 < m^2$) and the momentum
 transfer $|\bq|$ large enough,  one can expect that the
 additional two structure functions are
 small corrections, so that the  SIDIS cross section can  still be described by
  one, effective structure function $F_{2A}^{s.i.}(Q^2,x,p_2)$
 \cite{scopetta,ckk,simula,sarg}. Note, that in the considered SIDIS processes the
   final proton can originate from two different
   mechanisms, i) a nucleon from the target (different from
   the detected proton, i.e., the neutron in the deuteron case)
is struck by  the (highly) virtual photon producing the final hadron debris, while
   the detected proton acts merely as a spectator (the so-called spectator mechanism)
   ii) the highly excited quark-gluon system,
   produced by the interaction of the photon with an active proton
hadronizes into a final detected proton and a number of (undetected) mesons, which together
  with the nuclear remnants, form the undetected debris $X$ (the so called
    target fragmentation mechanism).


    The magnitude of the momentum of the
   proton from  the target fragmentation is similar to the momentum of protons resulting
   from the spectator mechanism, therefore we do not  consider here protons arising
   from current fragmentation  which have a much higher value of the momentum.

\subsection{The spectator mechanism}\label{sec:tri}
 As already mentioned,  according to the spectator mechanism,
  the deep-inelastic electromagnetic  process occurs on the $\it active$
nucleon, e.g. nucleon "1",  while  the second one (the  $ spectator$)
 recoils and is detected in coincidence with the scattered electron.
At high values of the  3-momentum transfer
the produced hadron debris propagates mostly  along the $\bf q$
direction and  re-interacts with  the spectator nucleon.  The wave function  of such a state
 can be written in the form
\be &&
\Psi_f (\{ \xi \},{\bf r}_X,{\bf r}_s ) =  \
\phi_{\beta_f}(\{ \xi\} )\psi_{p_X,p_s}({\bf r}_X,{\bf r}_s),
\label{eq14}
\ee
where ${\bf r}_X$ and ${\bf r}_s$  are the coordinates of the
center-of-mass of   system $X$ and the spectator nucleon, respectively,
$\{ \xi\} $ denotes the set of the internal coordinates of  system $X$,
described by the internal wave function
$\phi_{\beta_f}(\{ \xi\})$,  $\beta_f$ denoting  all  quantum numbers
 of the final state $X$;
the wave function  $\psi_{p_x,p_s}({\bf r}_x,{\bf r}_s)$  describes  the relative
motion of  system $X$ interacting elastically  with  the spectator $s$.
The matrix elements in  Eq. (\ref{eq5}) can  easily be  computed, provided
 the  contribution of the two-body  part of the deuteron electro-magnetic current
can be disregarded, which means that  the deuteron
current  can be  represented as a sum of electromagnetic
 currents of individual nucleons,
 $J_\mu (Q^2,X)=j_\mu^{N_1} + j_\mu^{N_2} $.
Introducing  complete sets of plane wave states
 $|{\bf k}_1{'},{\bf k}_2{'} \ra$ and $|{\bf k}_1,{\bf k}_2\ra$ in intermediate states,
 one obtains

\be&&
\la {\bf P}_D| j_\mu^N|\beta_f,{\bf P}_f={\bf p}_X+{\bf p}_2\ra=\nonumber \\&&
\sum\limits_{\beta,{\bf k}_1{'},{\bf k}_2{'}}
\sum\limits_{{\bf k}_1,{\bf k}_2}
\la {\bf P}_D|{\bf k}_1{'},{\bf k}_2{'}\ra
\la {\bf k}_1{'},{\bf k}_2{'}|j_\mu^{N_1}| \beta, {\bf k}_1,{\bf k}_2\ra
\la\beta, {\bf k}_1,{\bf k}_2|\ \beta_f,{\bf k}_X,{\bf k}_2
\ra =\nonumber\\&&
\int \frac{d^3 k}{(2\pi)^3} \psi_D({\bf k}_1)
\la {\bf k}_1|j_\mu^{N_1} (Q^2,p\cdot q)| \beta_f, {\bf k}_1+{\bf q}\ra
\psi_{ {\boldkappa}_f}({\bf q}/2-{\bf p}_2),
\label{eq15}
\ee
 where ${\boldkappa}_f=({\bf p}_X-{\bf p}_2)/2$.
 In Eq. (\ref{eq15}) the matrix element
 $\la {\bf k}_1|j_\mu^{N_1}(Q^2,k\cdot q)| \beta_f, {\bf k}_1+{\bf q}\ra$
 describes the electromagnetic transition from a moving nucleon
 in the initial  state to a final hadronic system $X$ in a quantum state $\beta_f$.
The  sum over all  final state $\beta_f$ of the
  square of this matrix element times the corresponding energy conservation
  $\delta$-function defines  the deep inelastic
  nucleon hadronic tensor for a moving nucleon.

\subsubsection{The PWIA}\label{subsect:triA}
In PWIA, $\gamma^*$ interacts with a quark of the neutron, a nucleon debris is formed and
the proton recoils without interacting with the debris.
The relative motion debris-proton is thus described by a plane wave
\be &&
\psi_{\boldkappa_f}({\bf q}/2-{\bf k}_2)\sim
(2\pi)^3\delta^{(3)} ({\bf q}/2-{\bf k}_2- \boldkappa_f)=
(2\pi)^3\delta^{(3)} ({\bf k}_2-{\bf p}_2)
\label{eq16}
\ee
and the well known result
\be
&&
\frac{d\sigma_{sp}^{PWIA}}{dx dQ^2\ d\bp_2}
=
K(x,Q^2,p_2)\, n_D(|{\bf k}_1|) \, F_2^{N/A}(x,p_2),
\label{crossPWIA}
\ee
is obtained, where
 the kinematical factor $K(x,Q^2,p_s)$,
in the Bjorken limit is (see, e.g. ref. \cite{scopetta})
\beq
K(x,Q^2,p_2)=\,\frac{4\pi\alpha_{em}^2}{xQ^4}
\bigg[1-y+\frac{y^2}{2}\bigg],
\label{kfactor}
\eeq
 $F_2^{N/A}(x,p_2) = 2x F_1^{N/A}(x,p_2)$ is  the DIS structure
function of  the  active nucleon,
and $n_D$  the momentum distribution of the hit nucleon with $\bk_1=-\bp_2$, {\it viz}
\be
n_D(|{\bf k}_1|)=\frac13\frac{1}{(2\pi)^3} \sum\limits_{{\cal M}_D} \left |\int d^3 r
 \Psi_{{1,\cal
M}_D}( {\bf r})\exp(-i{\bf k}_1{\bf  r}/2) \right|^2.
\label{dismom}
\ee

\subsubsection{FSI}\label{subsect:triB}
  Consider now  FSI effects  within  the   kinematics
when  the   momentum of the spectator is low
 and  the momentum transfer is large enough, so that
 the rescattering process of the fast system $X$ off the spectator
 nucleon could be considered as a high-energy soft hadronic interaction.
 In this case the momentum of the detected spectator ${\bf p}_2$
 only slightly differs from the  momentum ${\bf k}_2$ before rescattering, so that
  in
 the matrix element
  $\la -{\bf k}_2|j_\mu^{N_1} (Q^2,p\cdot q)|\beta_f, -{\bf k}_2+{\bf q}\ra$
   one can take  ${\bf k}_2\sim  {\bf p}_2$,  obtaining
in co-ordinate space
\be
&&
\la {\bf P}_D| j_\mu^N|{\bf P}_f\ra\cong j_{\mu}^N(Q^2,x, {\bf p}_2)
\int d^3 r
 \psi_D({\bf r}) \psi_{\boldkappa_f}^+({\bf r})\exp(i{\bf r q}/2).
\label{eq23}
\ee
 The cross section then becomes

\be
&&
\frac{d\sigma_{sp}^{FSI}}{dx dQ^2\ d\bp_2}
=
K(x,Q^2,p_2)\, n_D^{FSI}({\bf p}_2,{\bf q}) \, F_2^{N/A}(x,p_2),
\label{crossfsi}
\ee
where
\be
n_D^{FSI}({\bf p}_2,{\bf q})=\frac13\frac{1}{(2\pi)^3} \sum\limits_{{\cal
M}_D} \left |\int d^3 r
 \Psi_{{1,\cal M}_D}( {\bf r}) \psi_{\boldkappa_f}^+({\bf r})\exp(i{\bf r q}/2) \right|^2,
\label{dismomfsi1}
\ee
\noindent is the distorted momentum distribution, which coincides with the
 momentum distribution of the hit nucleon
(Eq.~(\ref{dismom}))
 when $\psi_{{\boldkappa}_f}^+({\bf r})\sim \exp(-i{\boldkappa}_f\bf r)$ .

In our  case, when the relative momentum  $\boldkappa_f \sim  {\bf q}$
 is rather large and the rescattering processes occur with low momentum transfers, the
 wave function $\psi_{{\boldkappa}_f}^+({\bf r})$
  can be replaced by its eikonal form describing
  the propagation of the
 nucleon  debris formed after  $\gamma^*$ absorption by a target quark, followed  by
  hadronization processes and interactions
 of the newly produced hadrons with the spectator nucleon. This series of soft
 interactions with the spectator can be characterized by an effective cross
 section  $\sigma_{eff}(z,Q^2,x)$ \cite{ckk,ciokop}
  depending upon time (or the distance $z$ traveled  by the system $X$).
 Thus the distorted  nucleon momentum distribution,  Eq.~(\ref{dismomfsi1}), becomes

\begin{equation}
 n_D^{FSI}( {\bf p}_s,{\bf q}) =
\frac13\frac{1}{(2\pi)^3} \sum\limits_{{\cal
M}_D} \left | \int\, d  {\bf r} \Psi_{{1,\cal
M}_D}( {\bf r}) S( {\bf r},{\bf q}) \chi_f^+\,\exp (-i
{\bf p}_s {\bf r}) \right |^2,
 \label{dismomfsi}
\end{equation}
where $\chi_f$ is the  spin function of the spectator nucleon and
$S( {\bf r},{\bf q})$   the $S$-matrix describing
the  final state interaction
between the debris and the  spectator,  {\it viz.}
\begin{equation}
S({\bf r},{\bf q}) = 1-\theta(z)\, \frac{\sigma_{eff}(z,Q^2,x)(1-i\alpha)}{4\pi b_0^2}\,
\exp(-b^2/2b_0^2).
 \label{gama}
\end{equation}

\noindent where the $z$ axis is directed along ${\bf q}$, i.e.
${\bf r} = z \displaystyle\frac{\bf q}{|{\bf q}|}+\bf b$.

\subsection{Target fragmentation}

\begin{figure}[!ht]
\centerline{\hspace{-.1cm}
\includegraphics[scale=0.4 ,angle=0]{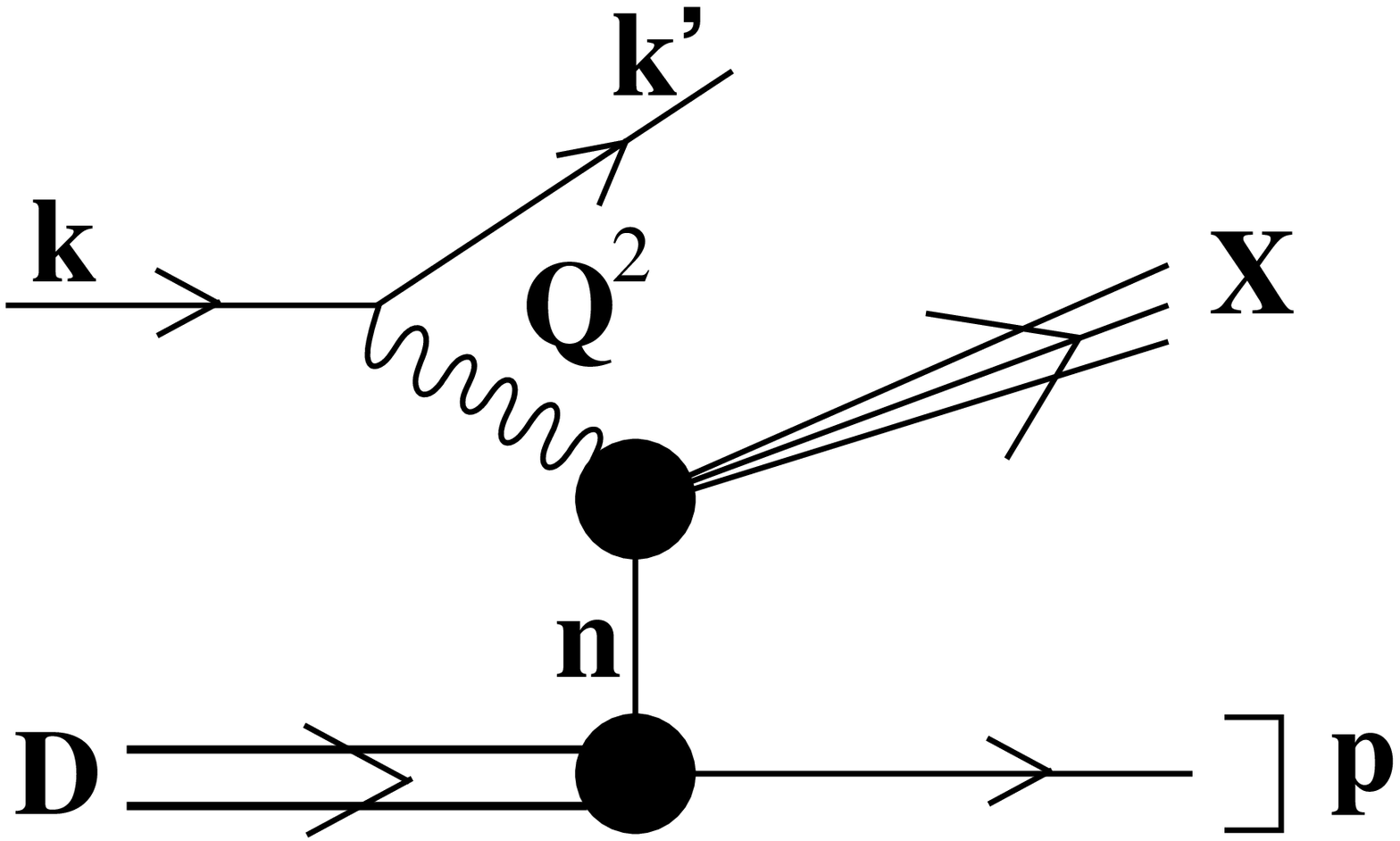}
 \hspace{3cm}
\includegraphics[scale=0.35,angle=0]{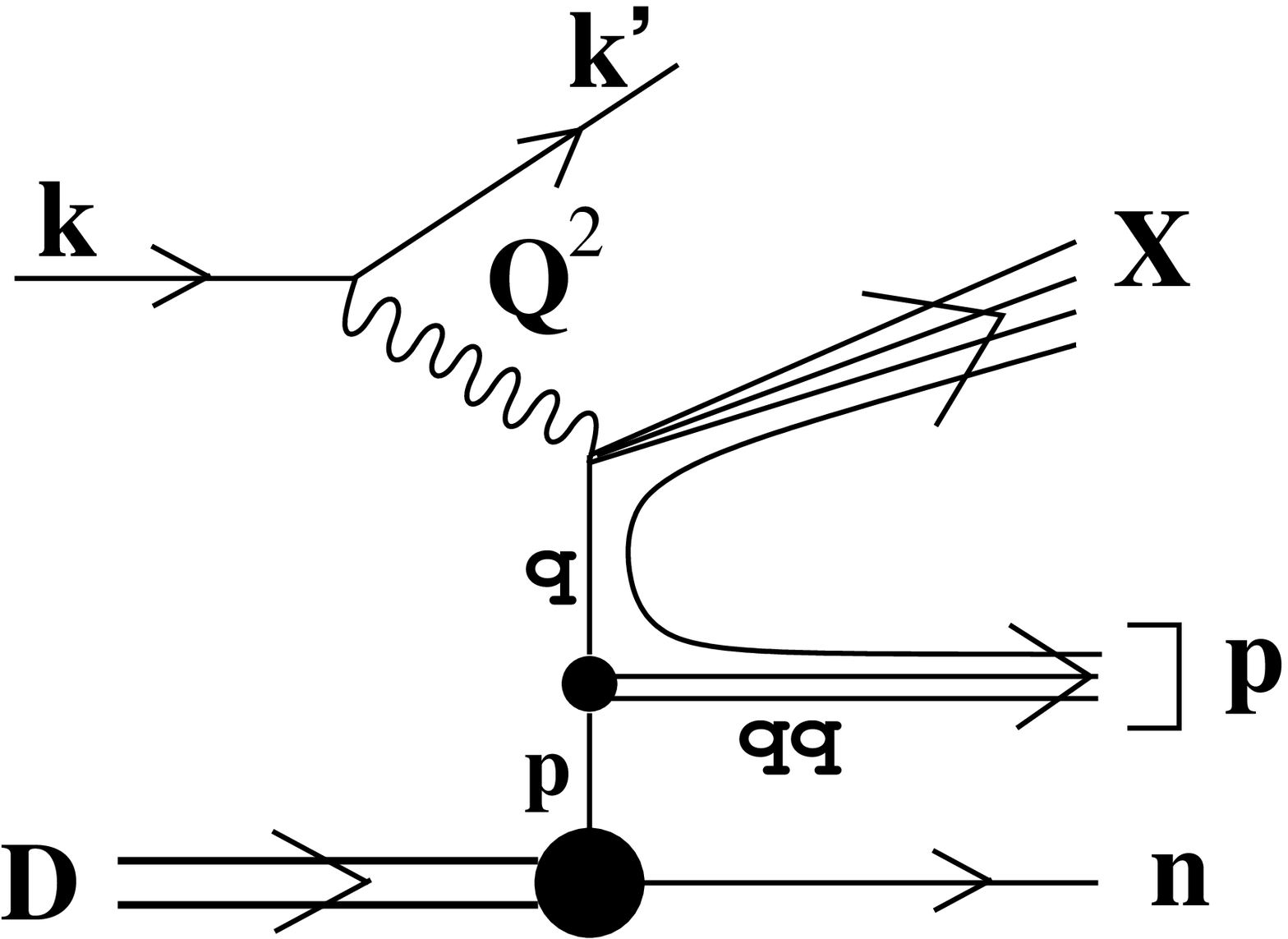}}
\caption{The Feynman diagrams of the process $D(e,e'p)X$  corresponding to the
spectator (left panel)  and to target fragmentation (right panel) mechanisms.}
\label{Dtarget}
\end{figure}

The target fragmentation mechanism is rather different from the spectator one
(c.f. Fig.~\ref{Dtarget}). However, by introducing
the notion of  fragmentation functions $H_{1,2}(Q^2,x,p_2)$~\cite{Ffunction},
 the theoretical analysis
 of both target fragmentation and  spectator mechanisms  becomes
 similar and  a common theoretical framework can be used.
The only difference consists in replacing the nuclear DIS structure function
$F_{2A}^{s.i.}(Q^2,x,p_2)$ by the nuclear fragmentation function
$H_{2A}(Q^2,x,p_2)$.  Then  in the Bjorken limit the corresponding  cross section reads as
follows

\beq
\frac{d\sigma_{t.f.}}{dx dQ^2\ d\bp_{s}}
=
K(x,Q^2,p_2)  H_2^D\left(x,z_s,\ner{p}^2_{s\bot}\right)
\label{tf}
\eeq

\noindent where the kinematical factor $K(x,Q^2,p_2)$ is given by
  Eq.~(\ref{kfactor}).
In Eq.~(\ref{tf}) the deuteron target fragmentation structure function
$H_2^D\left(x,z_s,\ner{p}^2_{s\bot}\right)$ can be expressed as a
convolution of the nucleon
distribution  with the corresponding nucleon structure function
$H_2^N\left(x,z_s,\ner{p}^2_{s\bot}\right)$ as follows
\beq
H_2^D\left(x,z_s,\ner{p}^2_{s\bot}\right)\,=\,\int_{x+z_p}^{M_D/m_N} dz_1\,f(z_1)\,
H_2^N\left(\frac{x}{z_1},z_s\,\frac{1-x}{z_1-x},\ner{p}_{s\bot}^2\right)\, ,
\eeq
where
\beq
f(z_1)\,=\,2\pi\,m_N\,z_1\,\int_{p_{min}}^{\infty}\,d|\bp|\,|\bp|\,n_D(\bp)\,
\label{fz}
\eeq
with $z_s=(p_2 q)/M_N\nu$,  $z_p=z_s(1-x)$   and
$p_{min}=\left |[(m_Nz_1-M_D)^{2\phantom{^1}}-m_N^2]/[2(m_Nz_1-M_D)]\right|$.
The nucleon structure function $H_2^N(x,z_s,\ner{p}_{s\bot}^2)$ is
\beq
H_2^N\left(x,z_s,\ner{p}_{s\bot}^2\right)\,=\, x\,\frac{\rho(\ner{p}_{s\bot})}{E_h}\,z_s\,
\Big[ \sum_q e_q^2f_q(x) D^p_{qq}(z_s)\Big]\,,
\label{h2n}
\eeq
where $\rho(\ner{p}_{s\bot})$ is the transverse momentum distribution of the
nucleon, $f_q(x)$ is the parton distribution function,  and   $D^p_{qq}(z_s)$
is the diquark fragmentation function representing  the probability
to produce a
proton with energy  fraction $z_s$ from a diquark.
The nucleon  functions $H_2^N(x,z_s,\ner{p}_{s\bot}^2)$,
likewise the DIS structure functions  $F_2^N(x,Q^2)$, are well  known
experimentally and the  parametrized form  of  $\rho(\ner{p}_{s\bot})$
and $D^p_{qq}(z_s)$ in Eq.~(\ref{h2n}) can be found, e.g. in Refs. \cite{distrtr,barframaj}.

\subsection{Results for the deuteron target}
\begin{figure}[!ht]
\includegraphics[scale=1.7,angle=0]{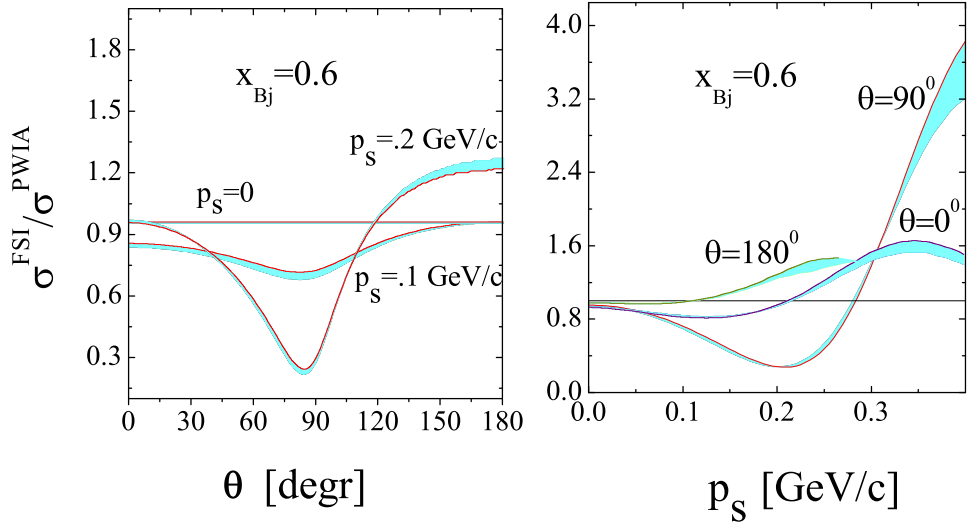}
\caption{An illustration of the role of FSI in $D(e,e'p)X$ processes
within the spectator mechanism. Left panel: angular dependence
of the ratio $\sigma^{FSI}/\sigma^{PWIA}$  at several fixed values of the
spectator momentum. Right panel: dependence of the ratio $\sigma^{FSI}/\sigma^{PWIA}$
upon the momentum of the spectator proton at parallel ($\theta=0^o$ and $\theta=180^o$)
and perpendicular ($\theta=90^o$) kinematics. Since the actual parametrizations
 of the nucleon DIS structure functions $F_2^2(x,Q^2)$ depend upon $Q^2$,
  in our calculations we put $Q^2=12 GeV^2/c^2$. The chosen
 kinematics is close to the one  planned in the future
experiments at JLAB at  $\CE \sim 12 GeV$. The shaded
 area is due to  the uncertainties in the input parameters for
 $\sigma_{eff}$ in Eq. (\ref{gama}).  }
\label{bonus12}
\end{figure}

\begin{figure}[!ht]
\includegraphics[scale=1.7,angle=0]{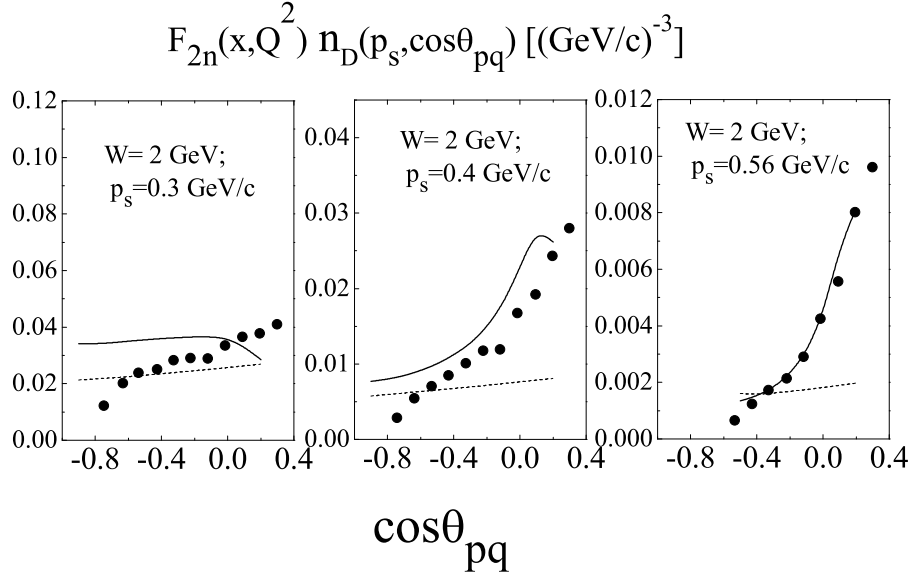}
\caption{The reduced cross section  (\ref{crossfsi}) (i.e. divided
by a proper kinematical factor),  representing, within the PWIA, the
 product of the  neutron DIS structure function
$F_{2n}(x/z_1,Q^2)$ and the deuteron momentum distribution $n_D(|\bp_s|)$, {\it  vs.}
the proton emission angle for different values of
$|\bp_s|$ and  fixed  value of the invariant mass of the debris $X$,
$W =\sqrt{(P_D-p_2+q)^2}$ and $Q^2=1.8\  GeV^2/c^2$. The dashed curves correspond
to PWIA results, whereas  the full ones include the effects of FSI.
Experimental data  from ref. \cite{KuhnPhysRev}. }
\label{bonus}
\end{figure}

  In order to analyze the kinematical conditions under which the effects of FSI
  are minimized
  or maximized,
  the ratio of the  PWIA cross section (Eq.~(\ref{crossPWIA}))
  to the cross section including FSI (Eq.~(\ref{crossfsi})) has been considered.
  Obviously,
  such a quantity reflects the effects of FSI in the distorted momentum distribution
  $n_D^{FSI}( {\bf p}_s,{\bf q}) $ ( Eq.~(\ref{dismomfsi})). The results of
  calculations are presented in Fig.~\ref{bonus12}, where
  the angular dependence (left panel) and the dependence
 upon the value of  the spectator momentum (right panel),  are shown at
  $x=0.6$. Kinematics has
  been chosen so as to correspond to the one considered  at the Jlab
  experiments at  $12 GeV$. The shaded area reflects the
  uncertainties  in the choice of the parameters for $\sigma_{eff}$ \cite{ciokop},
  $b_0$ and $\alpha$ in Eq.~(\ref{gama}).  It is clearly seen
  that at low momenta and at  parallel kinematics the effects of FSI are minimized, so that
  in this region the process $D(e,e'p)X$ could  be successfully used to extract
  the DIS structure function of a bound nucleon.
  Contrarily, at the perpendicular kinematics the FSI effects are rather important and
  essentially depend upon the process of hadronization of the quark-gluon debris.
   Therefore, in this region,  the
  processes $D(e,e'p)X$ can serve
  as a source of unique information about
  nonperturbative QCD mechanisms in DIS.
Actually, a  systematic experimental study of the processes $D(e,e'p)X$
has started at Jlab \cite{bonus} and first experimental data at initial
electron energy $E=5.765 GeV$ are already available \cite{KuhnPhysRev}.
It should be  however noted, that our approach corresponds to  the Bjorken limit,
so that a direct comparison with the  experimental data of ref. \cite{KuhnPhysRev}
requires the generalization  of our formulae to finite values of $Q^2$ and
$\nu$, when  the Callan-Gross relation is violated and a
proper modification of the hadronization mechanism is required.
Such an analysis will be presented in details  elsewhere \cite{nash_Kuhn_paper}.
In order to minimize the statistical errors, it is common in the literature to present
 the ratio of  the  experimental cross section to  a properly chosen
 kinematical factor, which, within the  PWIA spectator mechanism, corresponds
to a product of the neutron DIS structure function $F_2^n(x,Q^2)$ and
the deuteron momentum distribution (\ref{dismom}). In such a way, one expects that any
deviation of this ratio from the corresponding theoretical ratio,
would represent a measure of FSI effects.
In Fig.~\ref{bonus} we present the results of calculations
of such a reduced cross section obtained within the spectator
mechanism. The dashed curves  represent the  PWIA results, whereas  the solid ones
 correspond to the calculations with FSI  taken
into account. One can conclude from Fig.~\ref{bonus}, that
the spectator mechanism within the PWIA does not explain the data in the whole kinematical range,
whereas an
overall better agreement  can be achieved
when  FSI effects are taken into account.

\begin{figure}[!ht]      
\centerline{\hspace{2cm}
\includegraphics[scale=0.35,angle=-90]{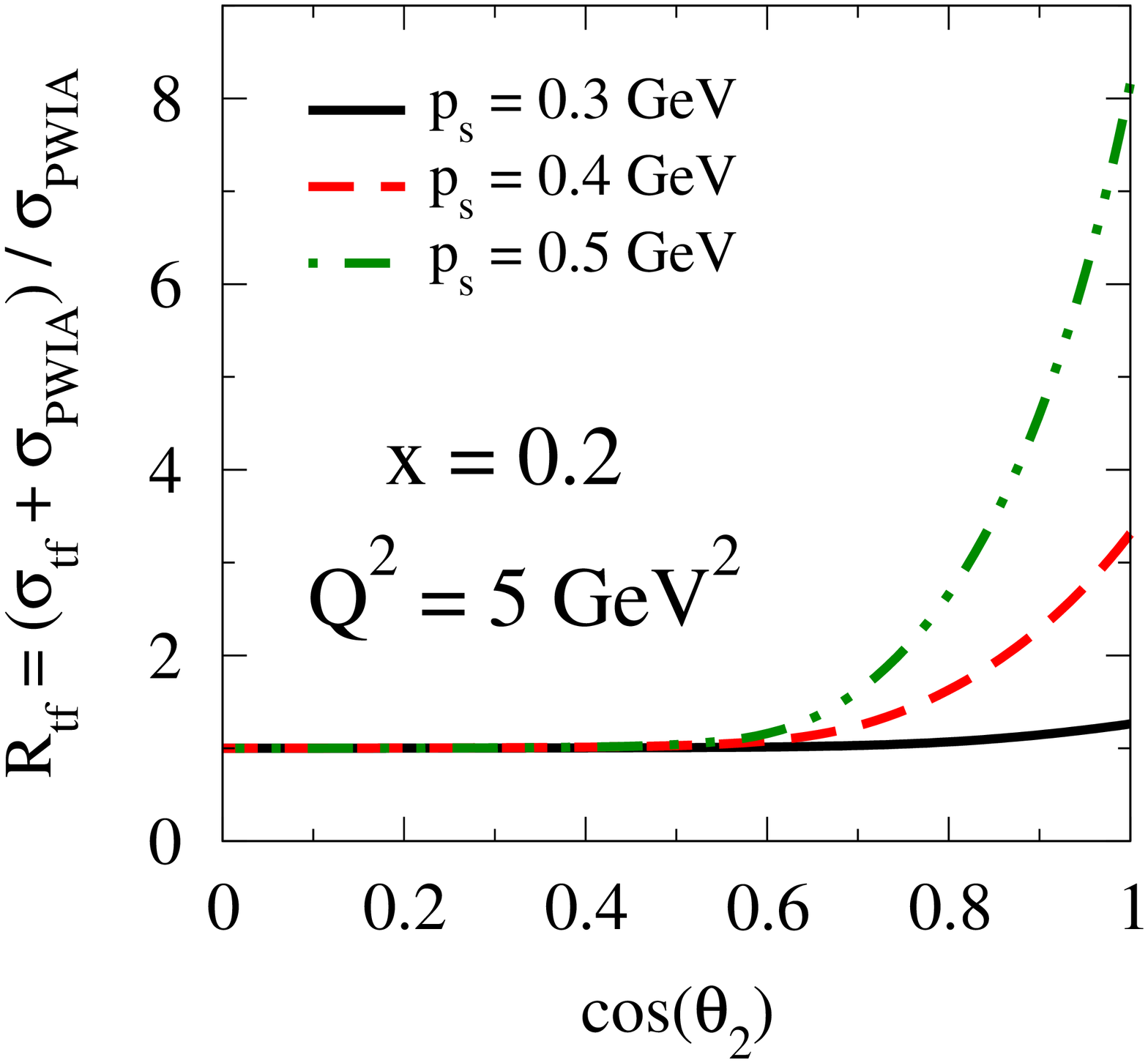}
\hspace{-2cm}
\includegraphics[scale=0.35,angle=-90]{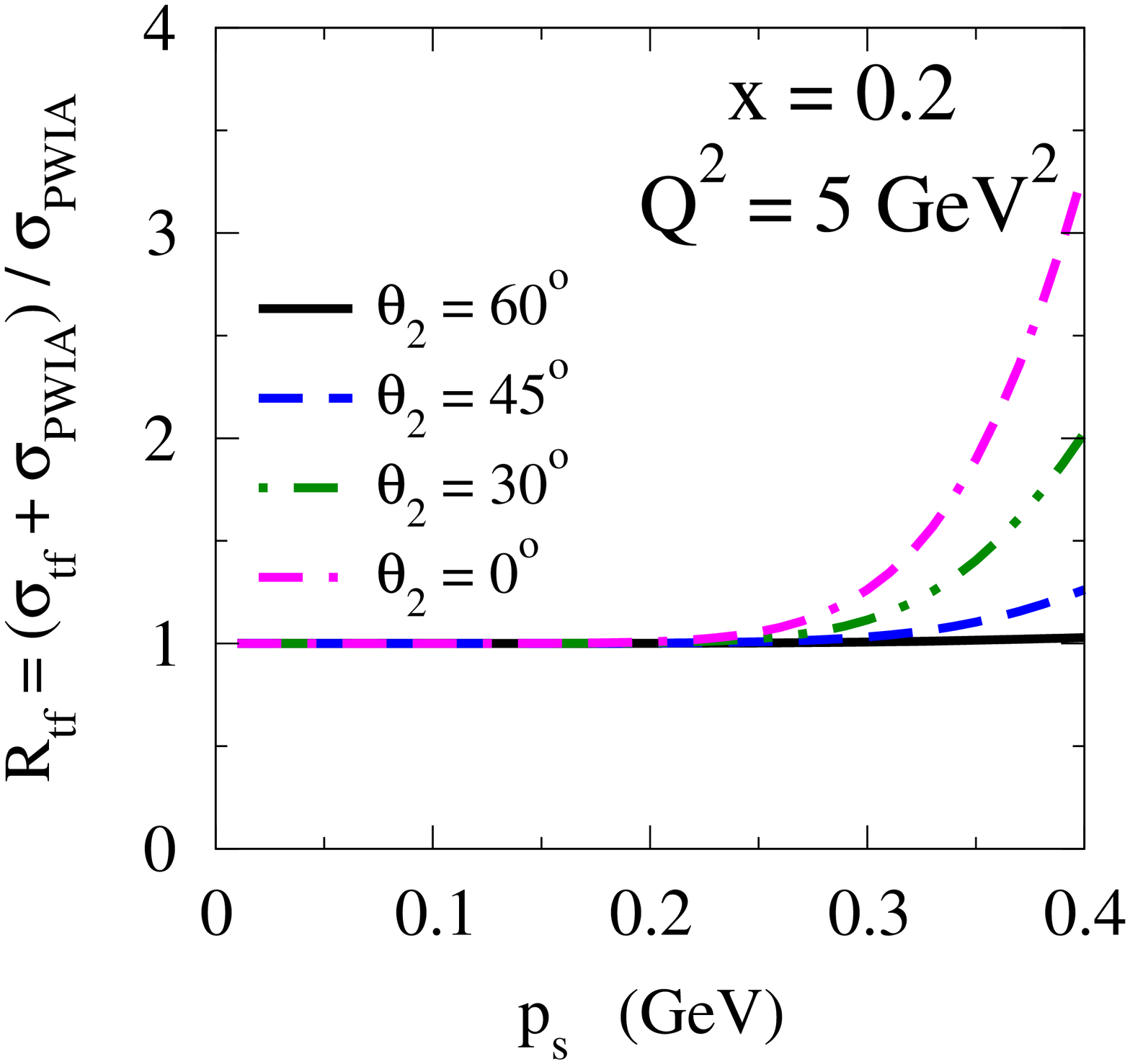}}
\caption{Nucleon emission by target fragmentation. The ratio
 $R_{tf}=(\sigma_{tf}+\sigma_{PWIA})/\sigma_{PWIA}$  is plotted {\it  vs.}
$cos\theta_2$ ($\theta_2$ is the proton emission angle)
and {\it  vs.}  $|\bp_s|$.}
\label{deut-frag-1}
\end{figure}

In order to estimate the role of the target fragmentation mechanism,
we have calculated the  ratio
\beq
R_{tf}\,=\,\frac{d\sigma_{tf}+d\sigma_{sp}^{PWIA}}{d\sigma_{sp}^{PWIA}}\, ,
\eeq
which, obviously,  characterizes the relative contribution of the
fragmentation cross section  to the total cross section.
In our calculations
the transverse hadron momentum distribution has been parametrized in the form
\cite{distrtr}
\beq
\rho(\ner{p}_{s\bot}) =
\frac{\beta}{\pi}\, \exp(-\beta\ner{p}_{s\bot}^2)\,,
\eeq
while the fragmentation function $D_{qq}$ has been  taken from ref. \cite{barframaj}.
The results of  calculations are shown in Fig. \ref{deut-frag-1},
where $R_{tf}$ is presented as a function of the emission angle of the detected proton
at several fixed values of the momentum (left panel) and the spectator momentum
at  fixed emission
angles (right panel).
As expected, the fragmentation mechanism contributes only in a very narrow forward
direction
and for large values of the spectator momentum.

\section{complex nuclei}
\subsection{The spectator mechanism within the PWIA}
In this section  our approach is generalized to complex nuclei
in the same way as it has been done in ref. \cite{ciosim2}.
The only essential difference with respect
to the formalism used in those papers is our consideration of the FSI of the quark-gluon
debris with the spectator  nucleons.
The basic nuclear ingredient in these calculations is the two-nucleon spectral function,
which has been chosen according to the nucleon-correlation model.
  In the simplest version of this model,
 the Center-of-Mass motion of correlated NN pair is disregarded ~\cite{FS}, whereas
 in the  extended  2NC model~\cite{ciosim} it is taken into account.

A calculation of the PWIA diagram  of Fig. \ref{fig6}(a)  yields
\beq
\frac{d\sigma}{dxdQ^2d\bp_2}\,=\,\frac{4\pi\alpha_{em}^2}{xQ^4}
\bigg[1-y+\frac{y^2}{2}\bigg]F_2^A(x,\ner{p}_2),
\label{compl}
\eeq
 where the nuclear structure function $F_2^A(x,p_2)$ is defined via the
 following convolution integral ~\cite{ciosim}

\beqy
F_2^A(x,p_2)&=&\int_x^{M_A/m_N-z_2}
\hspace{-0.5cm}dz_1\,z_1\,F_2^{N}(\frac{x}{z_1})
\int d\Vec{k}_{cm}\,dE^{(2)}\,S(\Vec{k}_{cm}-\Vec{k}_2,\Vec{k}_2,E^{(2)})\times\nonumber\\
&\times&m_N\delta(M_A-m_N(z_1+z_2)-M_{A-2}^fz_{A-2})
\label{qualesara}
\eeqy
where $\Vec{k}_{cm}=\Vec{k}_1 + \Vec{k}_2=-{\bf P}_{A-2}$,
$F_2^N({x}/{z_1})$ is the structure function of
the hit nucleon,
$z_1=(k_1q)/m_N\nu$, $z_2=[(m_N^2+{\bf p}_2^2)^{1/2}  - |{\bf p}_2|\cos\theta_2]/m_N$, and
$z_{A-2}=[( (M_{A-2}^f)^2+{\bf k}_{cm}^2 )^{1/2} + ({\bf k}\bq)/|\bq| ]/M_{A-2}^f$ are
 the light-cone momentum fractions of the hit nucleon, the detected nucleon and the recoiling
 spectator nucleus $A-2$,
 respectively,  $E^{(2)}$  denotes the two nucleon removal energy
 and the two nucleon spectral function
 $S(\Vec{k}_{1}=\Vec{k}_{cm}-\Vec{k}_{2}, \Vec{k}_{2},E^{(2)})$
is given by
\beq
\label{2ncex}
S(\Vec{k}_{1},\Vec{k}_{2},E^{(2)})\,=\,n_{rel}(|\Vec{k}_{1}-\Vec{k}_{2}|/2)
\,n_{cm}(|\Vec{k}_{1}+\Vec{k}_{2}|)\,\delta(E^{(2)}-E_{th}^{(2)})
\label{spfun}
\eeq
where $E_{th}^{(2)}$ is the two nucleon emission threshold.

\subsection{The Spectator Mechanism with FSI}
\begin{figure}      
\begin{center}
\includegraphics[width=4cm,height=3.9cm]{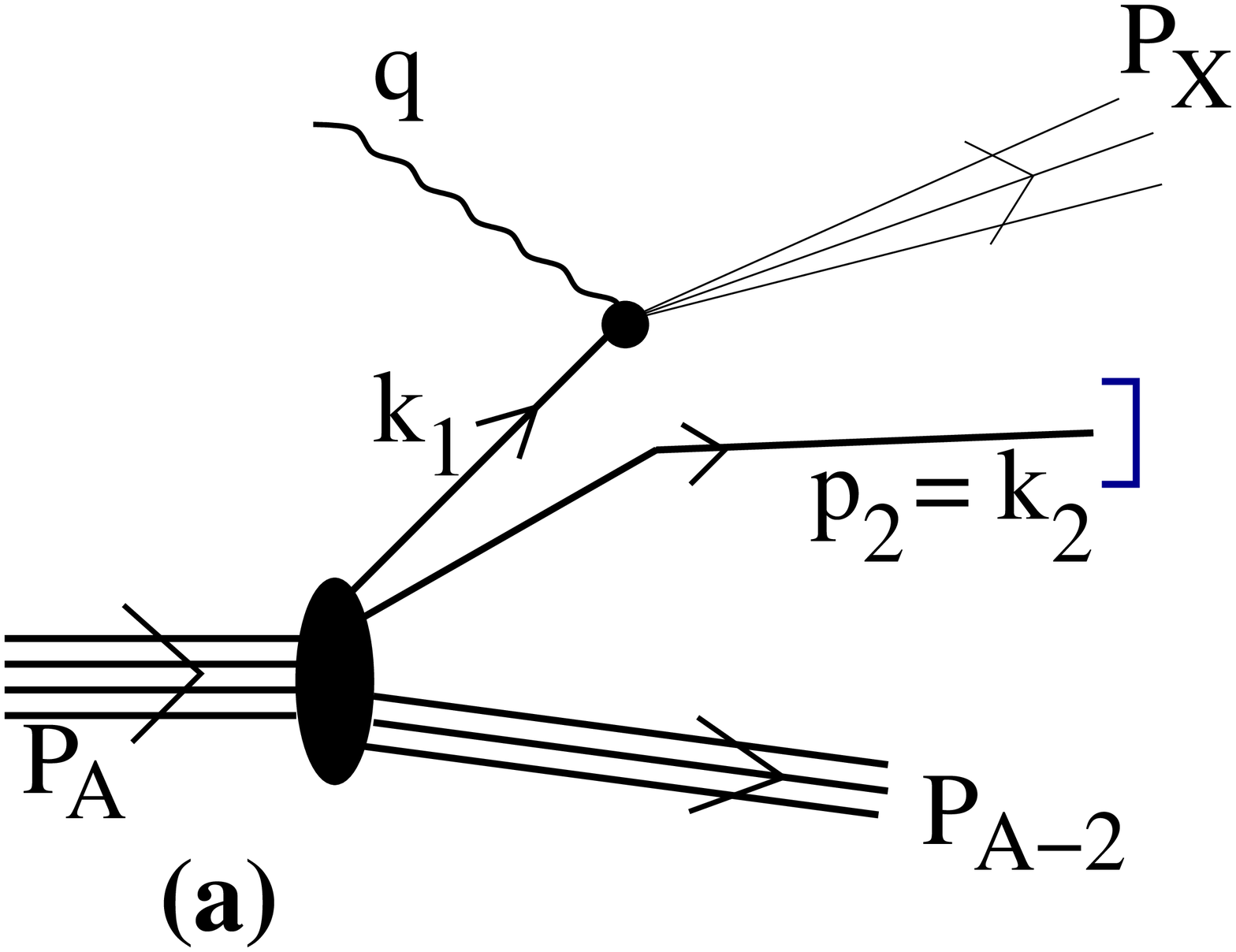}
\hspace{0.7cm}
\includegraphics[width=4cm,height=3.7cm]{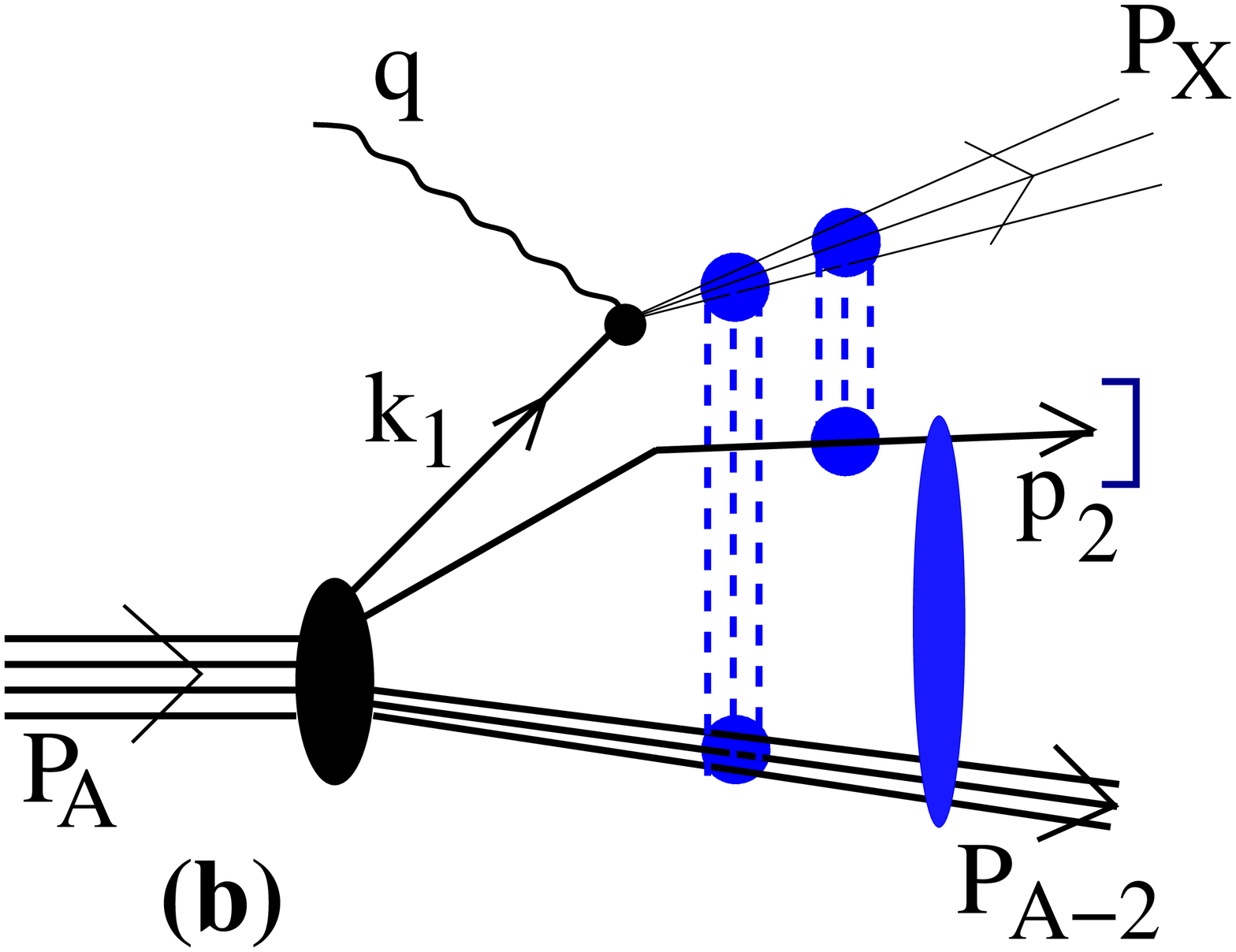}
\hspace{0.7cm}
\includegraphics[width=4cm,height=3.5cm]{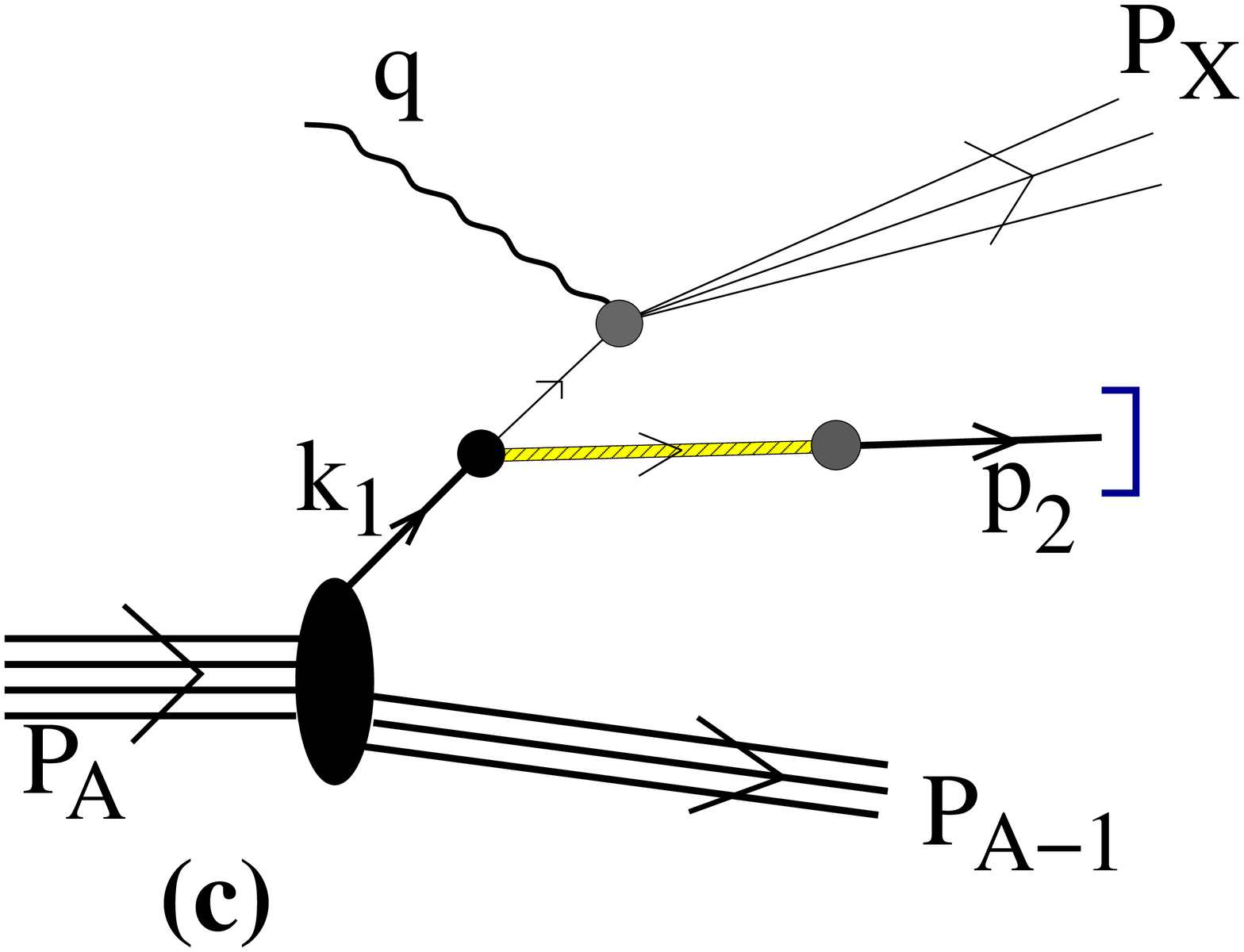}
\vskip -0.1cm
\caption{Proton production in   $A(e,e'p)X$ processes: (a)
spectator mechanism within the PWIA; (b)
various contribution to the FSI within the
spectator mechanism;  (c)  proton production from target fragmentation.}
\label{fig6}
\end{center}
\end{figure}

\n
Contrarily to the deuteron case, the effects of the FSI in complex nuclei
are much more complicated to treat  since in this case  the structure of the Spectral
Function (\ref{spfun})   implies that $(A-2)$ is in the ground state;
 in this case, after $\gamma^*$ absorption,  the final state consists
 of at least three different interacting  systems
 (c.f. Fig. ~\ref{fig6}(b) and ~\ref{fig6}(c)):
the undetected debris $X$, the undetected
$A-2$ nucleus and, eventually, the detected proton $p_2$.
Correspondingly, the FSI can formally be divided into three
classes~\cite{noi}, namely:
i) the FSI of the quark-gluon debris   with the spectator $A-2$
system;
ii) the interaction of the
 recoiling nucleon  with the $A-2$ system;
 iii) the interaction of  the  debris with the recoiling proton.

\subsubsection{FSI of the debris with the spectator  nucleons}
The FSI of the debris and $A-2$ system is treated in the same  way as in the deuteron case:
the quark-gluon debris rescatters off the recoiling proton and the $A-2$ spectator system.
Then in Eq. (\ref{qualesara}) the spectral function
$S(\bp_1,\bp_2,E^{(2)})$
 has to be replaced by the \textit{Distorted
spectral function}, which can be written in the following way
\beqy
S_{N_{1}N_{2}}^{D}(\bp_{1},\bp_{2},E^{(2)})&=&
\sum_{f}\,|T_{fi}|^{2}\,\delta(E^{(2)}-E_{th}^{(2)})=\nonumber\\
&=&\sum_{f}\,\left|\langle\bp_{X},\bp_{2},\Psi_{A-2}^{f},
\hat{S}_{FSI}\,|\,\bq,\Psi_{A}^{0}\rangle\right|^{2}\,\delta(E^{(2)}-E_{th}^{(2)}),
\eeqy
where $\hat{S}_{FSI}$ is the FSI operator and $T_{fi}$ is the transition matrix element of
the process
having the following form
\beqy
T_{fi}&=&\frac{1}{(2\pi)^6}\,\int \prod_{i=1}^{A}d\br_{i}\,
e^{-i\bp_{X}\cdot\br_{1}}\,e^{i\bq\cdot\br_{1}}
\,e^{-i\bp_{2}\cdot\br_{2}}\,\times\nonumber\\
&&\hspace{2cm}\times\,\Psi_{A-2}^{*f}(\br_{3},
\dots,\br_{A})\,\hat{S}_{FSI}^{+}(\br_{1},\dots,\br_{A})
\,\Psi_{A}^{0}(\br_{1},\dots,\br_{A})\,.
\eeqy
According to our  classification of the FSI effects,
the operator $\hat{S}_{FSI}$ will read as follows
\beq
\label{fsiop}
\hat{S}_{FSI}(\ner{r}_{1},\ner{r}_{2},\dots,\ner{r}_{A})\,=\,D_{p_{2}}(\ner{r}_{2})
\,G(\ner{r}_1,\ner{r}_2)\prod_{i=3}^{A}G(\ner{r}_1,\ner{r}_i)\,,
\eeq
where $D_{p_{2}}(\ner{r}_{2})$ and $G(\ner{r}_1,\ner{r}_2)$
take into account the interaction of the slow recoiling proton
with $A-2$ and with the fast nucleon debris, whereas $\prod_{i=3}^{A}G(\ner{r}_1,\ner{r}_i)$
takes into
account the interaction of
the nucleon debris with $A-2$.
Using momentum conservation $\ner{p}_{X}=\ner{q}-\ner{p}_{2}-\ner{P}_{A-2}$,
the transition matrix element of the process $A(e, e' p)X$ becomes:
\beqy
T_{fi}&=&\frac{1}{(2\pi)^6}\int \prod_{i=1}^{A}d\ner{r}_{i}\,
e^{i(\ner{P}_{A-2}+\ner{p}_{2})\cdot\ner{r}_{1}}
e^{-i\ner{p}_{2}\cdot\ner{r}_{2}}\times\nonumber\\
&&\times\,\Psi_{A-2}^{*f}(\ner{r}_{3},
\dots,\ner{r}_{A})\hat{S}_{FSI}^{+}(\ner{r}_{1},\dots,\ner{r}_{A})
\Psi_{A}^{0}(\ner{r}_{1},\dots,\ner{r}_{A})\,=\nonumber\\
&=&\frac{1}{(2\pi)^6}\int d\ner{r}_{1}d\ner{r}_{2}
\,e^{i(\ner{P}_{A-2}+\ner{p}_{2})\cdot\ner{r}_{1}}e^{-i\ner{p}_{2}\cdot\ner{r}_{2}}
\,I^{FSI}(\ner{r}_{1}, \ner{r}_{2})\,,
\eeqy
where
\beq
I^{FSI}(\ner{r}_{1}, \ner{r}_{2})\,=\,\int \prod_{i=3}^{A}d\ner{r}_{i}\,
\Psi_{A-2}^{*f}(\ner{r}_{3},\dots,\ner{r}_{A})
\hat{S}_{FSI}^{+}(\ner{r}_{1},\dots,\ner{r}_{A})
\Psi_{A}^{0}(\ner{r}_{1},\dots,\ner{r}_{A})
\label{twobodyphi}
\eeq
is the distorted two body overlap integral.

 We reiterate,  that in the present approach we consider protons
 with relatively large, at the average  fermi motion scale, momenta
  which originate from  a short-range correlated pair in the parent nucleus.
Then for such kinematics the nuclear wave function  can be
written as follows~\cite{ciosim}
\beq
\Psi_{A}^{0}(\ner{r}_{1},\dots,\ner{r}_{A})
\,=\, \sum_{\alpha\beta}
\Phi_{\alpha}(\ner{r}_1, \ner{r}_2)\otimes\Psi_{A-2}^{\beta}(\ner{r}_{3},\dots,\ner{r}_{A})\,,
\label{2NCwf}
\eeq
where $\Phi_{\alpha}(\ner{r}_1, \ner{r}_2)$ and $\Psi_{A-2}^{\beta}(\ner{r}_{3},\dots,
\ner{r}_{A})$
describe the correlated pair and the $A-2$ remnants, respectively.  In Eq.~(\ref{2NCwf})
the symbol $\otimes$ is used for a short-hand notation of the corresponding
Clebsh-Gordon coefficients. The wave function of the
correlated pair can be expanded over a complete set of wave functions describing the
intrinsic state of the pair and its motion   relative to the
$A-2$ kernel, viz.
\beq
\Phi_{\alpha}(\ner{r}_{1}, \ner{r}_{2})\,=\,\sum_{mn}\,c_{mn}\,\phi_m(\ner{r})\chi_n(\ner{R})\,,
\eeq
where $\ner{R}=\frac{1}{2}(\ner{r}_1+\ner{r}_2)$ and $\ner{r}=\ner{r}_1-\ner{r}_2$ are the
center of mass
and relative coordinate of the pair. As mentioned, in
the 2NC model is assumed that the correlated pair carries out the most part
of the nuclear momentum, while the
 momentum of the relative motion of the pair and $A-2$ nucleus is small~\cite{ciosim}.
 This allows one to treat the  CM motion   in its   lowest $^1S_0$ quantum state
 (in what follows denoted, for the sake of
 brevity, as $os$-state).
We can therefore write
\beq
\Phi_{\alpha}(\ner{r}_{1}, \ner{r}_{2})\,\simeq\,\chi_{os}(\ner{R})\,\sum_{m}c_{mo}
\,\phi_m(\ner{r})\,=\,
\chi_{os}(\ner{R})\,\varphi(\ner{r})\,\equiv\,\phi(\ner{r}_{1}, \ner{r}_{2})\,,
\label{f12}
\eeq
with
\beq
\varphi(\ner{r})\,=\,\sum_{m}\,c_{mo}\,\phi_m(\ner{r})
\eeq
Finally we have
\beq
\Psi_{A}^{0}(\ner{r}_{1},\dots,\ner{r}_{A})\,\simeq\,\chi_{os}(\ner{R})\,\varphi(\ner{r})
\,\Psi_{A-2}^{0}(\ner{r}_{3},\dots,\ner{r}_{A})\,.
\eeq
Placing this expression in Eq. (\ref{twobodyphi}) we get:
\beq
I^{FSI}(\ner{r}_1,\ner{r}_2)\,=\,\int \prod_{i=3}^{A}d\ner{r}_{i}\,
\chi_{os}(\ner{R})\,\varphi(\ner{r})\,\hat{S}^+_{FSI}(\ner{r}_{1},\dots,\ner{r}_{A})
\,|\Psi_{A-2}^{0}(\ner{r}_{3},\dots,\ner{r}_{A})|^2
\eeq
and, disregarding correlations in the $A-2$ system, one can write \cite{glau1,glau2}:
\beq
|\Psi_{A-2}^{0}(\ner{r}_{3},\dots,\ner{r}_{A})|^2\,\simeq\,\prod_{i=3}^A\rho(\ner{r}_i)\,,
\eeq
with $\int\rho(\ner{r}_i)d\ner{r}_i=1$, so that, eventually, the distorted overlap integral
becomes:
\beqy
I^{FSI}(\ner{r}_1,\ner{r}_2)&=&\int\prod_{i=3}^{A}d\ner{r}_{i}\,
\phi(\ner{r}_{1}, \ner{r}_{2})\prod_{i=3}^A\rho(\ner{r}_i)
D_{p_{2}}(\ner{r}_{2})\,G(\ner{r}_1,\ner{r}_2)\prod_{i=3}^{A}G(\ner{r}_1,\ner{r}_i)=
\nonumber\\
&=&\phi(\ner{r}_1, \ner{r}_2)\,G(\ner{r}_1,\ner{r}_2)
\,D_{p_{2}}(\ner{r}_{2})
\bigg[\int d\ner{r}\,\rho(\ner{r})\,G(\ner{r}_1,\ner{r})\bigg]^{A-2},
\eeqy
where $\phi(\ner{r}_{1}, \ner{r}_{2})\,=\,\chi_{os}(\ner{R})\,\varphi(\ner{r})$
(cf. Eq. \ref{f12}).
 In our calculations
the function $\phi(\ner{r}_{1}, \ner{r}_{2})$ has been chosen
phenomenologically in
such a way, that in PWIA to provide the same high momentum components
 of the two-nucleon spectral function, as reported in ref.~\cite{ciosim}.

The   explicit form of the operator
$G(\ner{r}_1,\ner{r}_i)$, describing the FSI of debris with spectators reads as \cite{ciokop}

\beq
\prod_{i=2}^A\,G(\ner{r}_1,\ner{r}_i)\,=\,\prod_{i=2}^A\,\Big[1-\theta(z_{i}-z_{1})\,
\Gamma(\ner{b}_{1}-\ner{b}_{i},z_{i}-z_{1})\Big]
\eeq
where $\ner{b}_i$ and $z_i$ are, the transverse and longitudinal components
of the coordinates of nucleon ``i'', and the function $\theta(z_{i}-z_{1})$ describes forward
debris propagation. The profile function in this case is determined by an effective cross
section,
which depends up on the distance, traveled by the string from the creation
to the corresponding hadronization points.
\beq
\Gamma(\ner{b}_{1}-\ner{b}_{i},z_{i}-z_{1})\,=\,\frac{(1-i\,\alpha)\,\,
\sigma_{eff}(z_{i}-z_{1})}
{4\,\pi\beta}\,e^{\displaystyle{-\frac{(\ner{b}_{1}-\ner{b}_{i})^{2}}{2\,\beta}}}
\eeq
where the cross section for the effective cross
section $\sigma_{eff}(z)$
 consists on  a sum of cross sections of  nucleon-nucleon and meson-nucleon
 interaction at the given point $z$,
$\sigma_{eff}(z)\,=\,\sigma_{tot}^{NN}\,+\,\sigma_{tot}^{\pi N}\,
\big[\,n_{M}(z)\,+\,n_{G}(z)\,\big]$, where
$n_{M}(z)$ and $n_{G}(z)$ are the effective numbers of mesons produced by
color string and gluon radiation, respectively.
 As demonstrated  in ref. \cite{ciokop1}, the considered  hadronization
 model for the  $\sigma_{eff}(z)$ provides a good description of grey track
 production in DIS off nuclei.
Now,  assuming that the single particle density
is normalized as $\int d\ner{r}\rho(\ner{r}) = 1$ and
is almost constant for complex nuclei, we can write:
\beqy
\hspace{-0.5cm}\bigg[\int d\ner{r}\,\rho(\ner{r})G(\ner{r}_1,\ner{r})\bigg]^{A-2}
\hspace{-0.5cm}&=&
\bigg[\int d\ner{r}\,\rho(\ner{r})-\int d\ner{b}\,\int^\infty_{z_1}dz\,\rho(\ner{b},z)
\,\Gamma(\ner{b}_1-\ner{b};z-z_1)\bigg]^{A-2}\simeq\nonumber\\
&\simeq&\bigg[1-\frac{1}{2}\int_{z_{1}}^{\infty} dz\,\rho(\ner{b}_1,z)
  \,\sigma_{eff}(z-z_1)\bigg]^{A-2}\,\simeq\nonumber\\
&\simeq&exp\left(-\frac{1}{2}\,A\int_{z_{1}}^{\infty} dz\,\rho(\ner{b}_1,z)
  \,\sigma_{eff}(z-z_1)\right)
\eeqy
and finally we can write the transition matrix element in the following way:
\beqy
T_{fi}&=&\frac{1}{(2\pi)^6}\int d\ner{r}_{1}\,d\ner{r}_{2}\,
e^{i\,(\ner{P}_{A-2}\,+\,\ner{p}_{2})\cdot\ner{r}_{1}}\,e^{-i\,\ner{p}_{2}\cdot\ner{r}_{2}}
\,\phi(\ner{r}_1, \ner{r}_2)\,\times\nonumber\\
&&\times\,G(\ner{r}_1,\ner{r}_2)\,D_{p_2}(\ner{r}_2)
\,exp\left(-\frac{1}{2}\,A\int_{z_{1}}^{\infty} dz\,\rho(\ner{b}_1,z)\,\sigma_{eff}(z-z_1)
\right)
\eeqy

\subsubsection{FSI of the  recoiling nucleon with the residual nucleus A-2}
\n
Here the effect of the interaction of the recoiling spectator nucleon with
the $A-2$ residual nucleus will be estimated.
In order to describe the motion of the spectator in the field of $A-2$
system, we use an Optical Potential approach,
also known as the conventional Distorted Wave Impulse Approximation
(DWIA). This approach is normally used to describe nucleon-nucleus interaction at low energies
\cite{boffi}. In the DWIA approach the interaction is considered as due to a mean field,
a distorting potential $V(x,y,z)$, which, in our framework, takes into account the
interaction between the residual nucleus and the emitted nucleon. The outgoing nucleon
plane wave becomes distorted as a consequence of the eikonal phase factor:
\beq
e^{-i\,\ner{p}_{2}\cdot\ner{r}_{2}}\longrightarrow e^{-i\,\ner{p}_{2}\cdot\ner{r}_{2}}
D_{p_{2}}(\ner{r}_{2}),
\eeq
where
\beq
D_{p_{2}}(\ner{r}_{2})\,=\,exp\left(i\,\frac{E_2}{p_{2}}\,
\int_{z_{2}}^{\infty}dz\,V(x_{2},y_{2},z)\right)\,.
\eeq
The optical potential $V(x,y,z)$ does not depend on the individual coordinates of the
spectator nucleons and embodies an effective description of how the nuclear medium influences
the wave function of the propagating nucleon.
We use an energy dependent optical potential:
\beq
V(\ner{r})\,=\,V(\ner{r})\,+\,i\,W(\ner{r})
\eeq
in which the real part $V$ represents elastic re-scattering and the imaginary part $W$,
reducing the proton flux, describes absorption processes.
The energy dependence of the potential is given by the well known choice
\beq
V(|\bp_2|)\,=\,-\rho\,\frac{v(i+\alpha)}{2}\,\sigma^{NN}_{tot}(|\bp_2|)\,,
\eeq
where $\rho$ is the nuclear density, $v$ the struck nucleon velocity, $\alpha$ the ratio of
the real to the
imaginary parts of the forward scattering amplitude $f_{NN}(0)$, and $\sigma^{NN}_{tot}(p)$
the total
nucleon nucleon scattering cross section, related to $f_{NN}(0)$ by the optical potential
$\sigma^{NN}_{tot}=\frac{4\pi}{p}Imf_{NN}(0)$. The parameters of
$\alpha$ and $\sigma^{NN}_{tot}$ have been taken from \cite{arn}.
When the energy of the propagating proton is small, each rescattering
causes a considerable loss of energy-momentum and the flux of the outgoing proton plane
 wave is
suppressed: this effect is modeled in the DWIA by the imaginary part of the potential.
Numerical calculations show  that the effect of the distortion
is mainly determined by the depth of the
imaginary part of the optical potential and
depends weakly on the real part, so we can
safely neglect it.

\subsection{The target fragmentation mechanism}
\noindent
Let us now consider  proton production  from the target fragmentation mechanism in which the
quark-gluon debris originates from current fragmentation and the proton arising
from target fragmentation
 (c.f.. Fig.~\ref{fig6}(c)).
 The corresponding cross section can be expressed in terms of  two
structure functions $H_1^A$ and $H_2^A$
 as follows

\beq
\frac{d\sigma^A}{dx\,dy\,d \bp_2} =
\frac{8\pi\alpha^2M \CE }{Q^4}\,\Bigg[x\,y^2\,H_1^A\left(x, z_s, \ner{p}_{s\bot}^2\right)
\,+\,(1-y)\,H_2^A\left(x,z_s, \ner{p}_{s\bot}^2\right)\Bigg]\,,
\eeq
where

\beq
z_s\,=\,\frac{(p_s q)/\nu}{m_N (1-x)}
\eeq
is the light-cone momentum fraction of the detected proton.
The structure functions  can be written  as
convolutions of the nucleon structure functions with the
familiar one-body nuclear spectral function $P(\bp,E_{r})$ as

\beq
H_1^A(x,z_s,\ner{p}_{s\bot}^2)\,=\,\int dz_1\,f(z_1)\,\frac{1}{z_1}\,
H_1^N\left(\frac{x}{z_1},z_s\,\frac{1-x}{z_1-x},\ner{p}_{s\bot}^2\right)\,,
\eeq
\beq
H_2^A\left(x,z_s,\ner{p}_{s\bot}^2\right)\,=\,\int dz_1\,f(z_1)\,
H_2^N\left(\frac{x}{z_1},z_s\,\frac{1-x}{z_1-x},\ner{p}_{s\bot}^2\right)\,,
\eeq
where $H_1^N$ and $H_2^N$ are the structure functions for the free nucleon and
$f(z_1)$  is given by
\beq
f(z_1)\,=\,\int d\bp \,dE_{r}\,P(\bp ,E_{r})\,z_1\,\delta\left(z_1-\frac{x}{x_N}\right)\,.
\label{effedue}
\eeq
 Within  the  the quark-parton model,  the nucleon structure functions
have the following form
\beq
H_1^N\left(x, z_s, \ner{p}_{s\bot}\right)\,=\,\frac{z_s}{E_2}
\,\rho(\ner{p}_{s\bot}^2)\,\frac{1}{2}\,\sum_q\,e_q^2\,f_q(x)\,
D_{qq}^h(z_s)
\eeq
\beq
H_2^N\left(x, z_s, \ner{p}_{s\bot}\right)\,=\,\frac{z_s}{E_2}\,\rho(\ner{p}_{s\bot}^2)
\,x\,\sum_q\,e_q^2\,f_q(x)\,D_{qq}^h(z_s)\,,
\eeq
where  $E_2$ is the energy of the detected proton and
$f_q(x)$  and $\rho(\ner{p}_{s\bot}^2)$
are the parallel and transverse momentum distributions, respectively.

\subsection{Results of calculations}
\begin{figure}[!ht]   
\centerline{\hspace{2cm}
\includegraphics[scale=0.33,angle=-90]{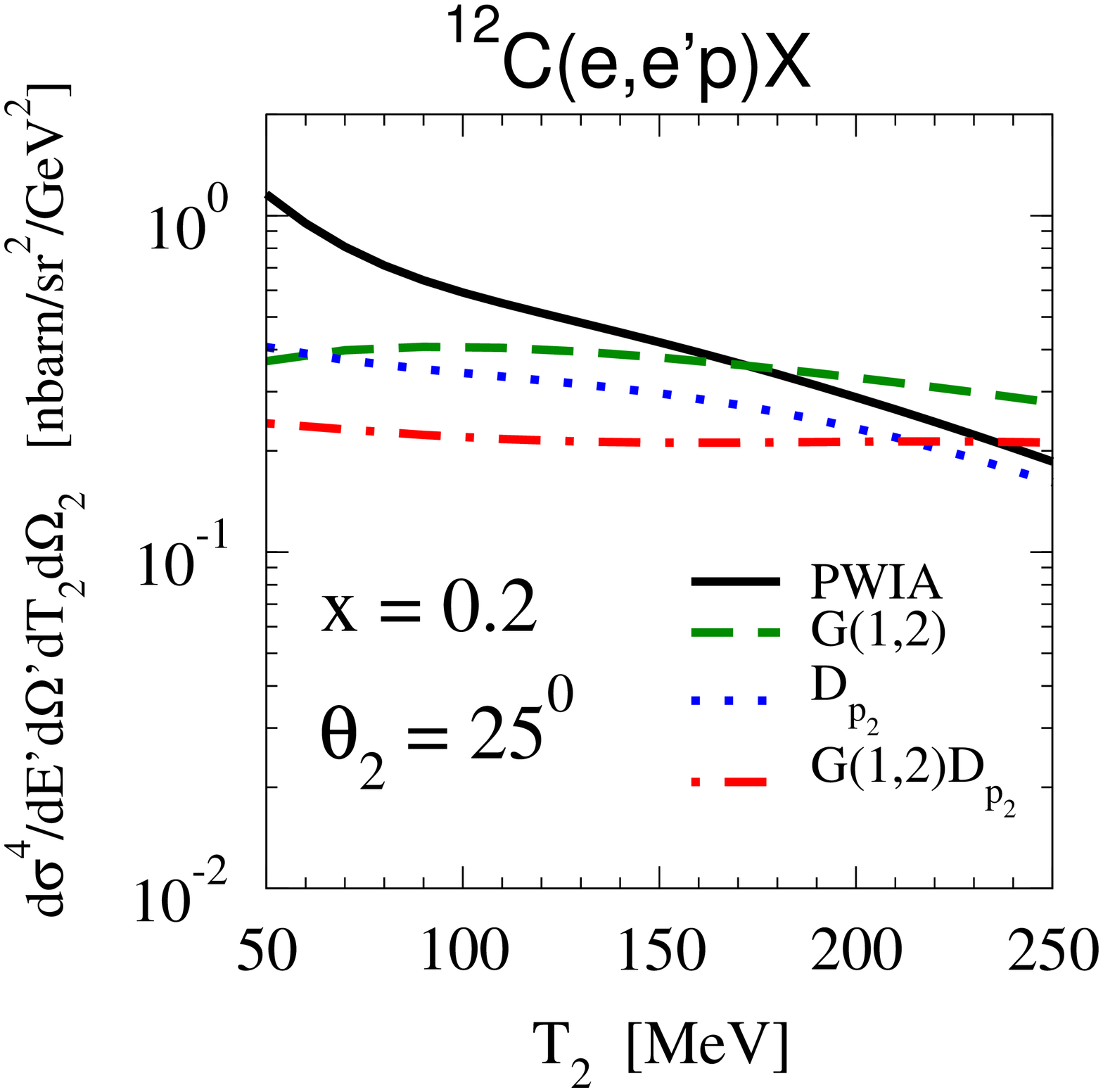}
\hspace{-2cm}
\includegraphics[scale=0.33,angle=-90]{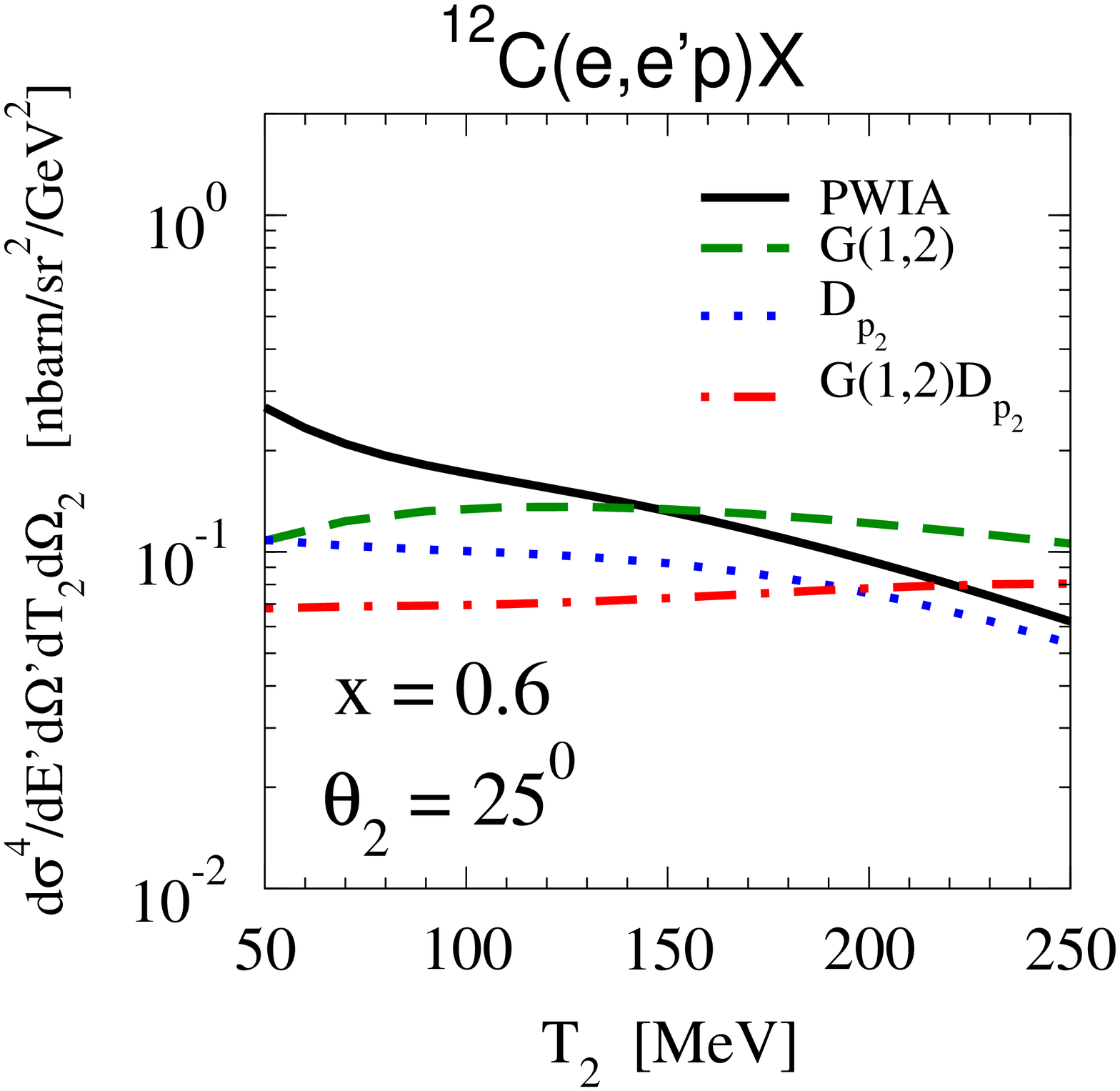}}
\caption{The SIDIS differential cross section for the process $^{12}C(e,e'p)X$ versus
the kinetic
energy $T_2$ of the detected proton, emitted forward at $\theta_2=25^0$ in correspondence of
 two values
of the Bjorken variable. Solid curve:  PWIA results;
dashed curve: PWIA plus  FSI of the debris $X$ with the
recoiling proton;  dotted curve: PWIA plus  FSI of the
proton with $A-2$ ; dot-dashed curve are the resulting
FSI effects from all the interactions.}
\label{contrsep}
\end{figure}

\begin{figure}[!hb]           
\centerline{\hspace{2cm}
\includegraphics[scale=0.33,angle=-90]{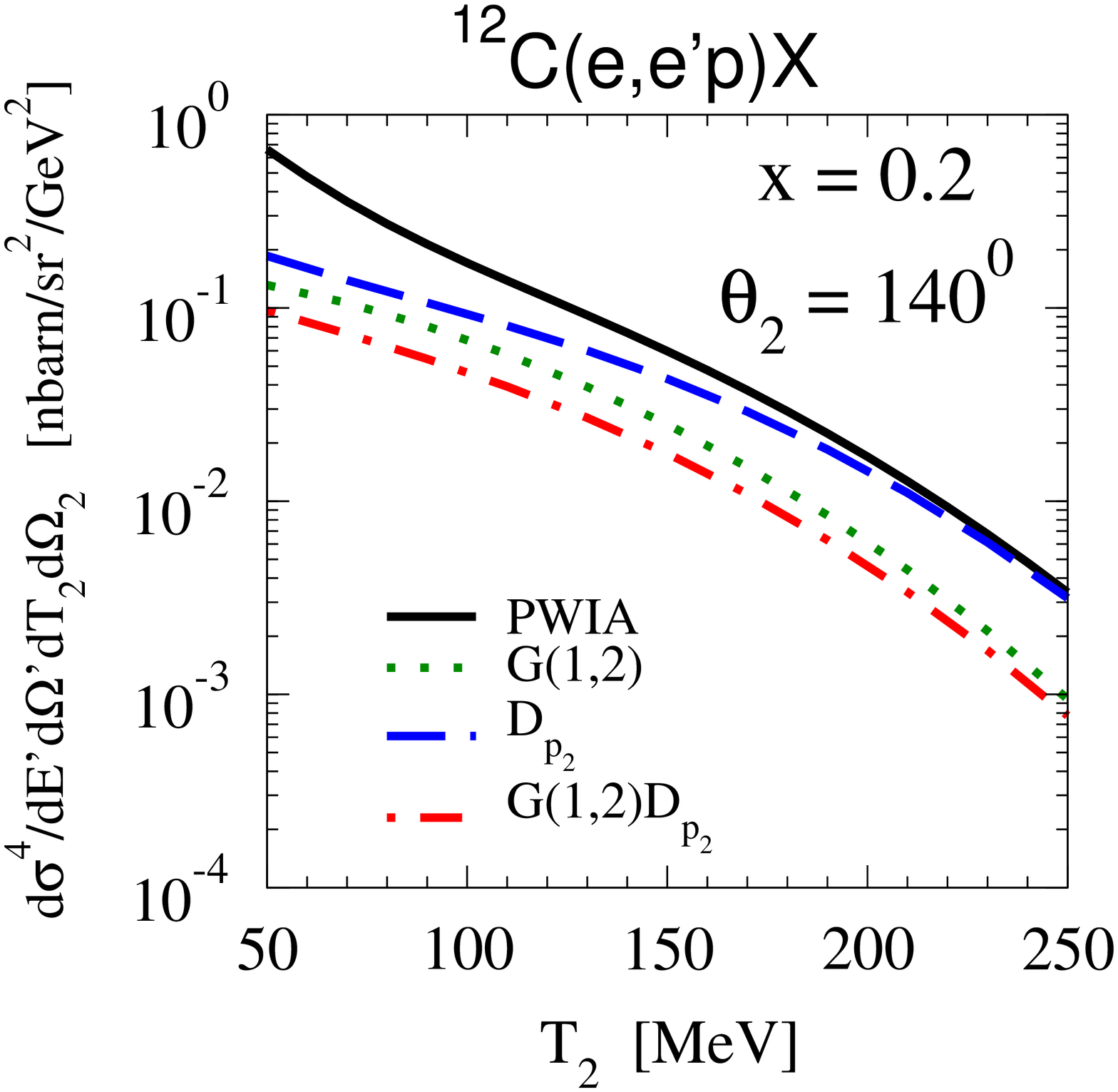}
\hspace{-2cm}
\includegraphics[scale=0.33,angle=-90]{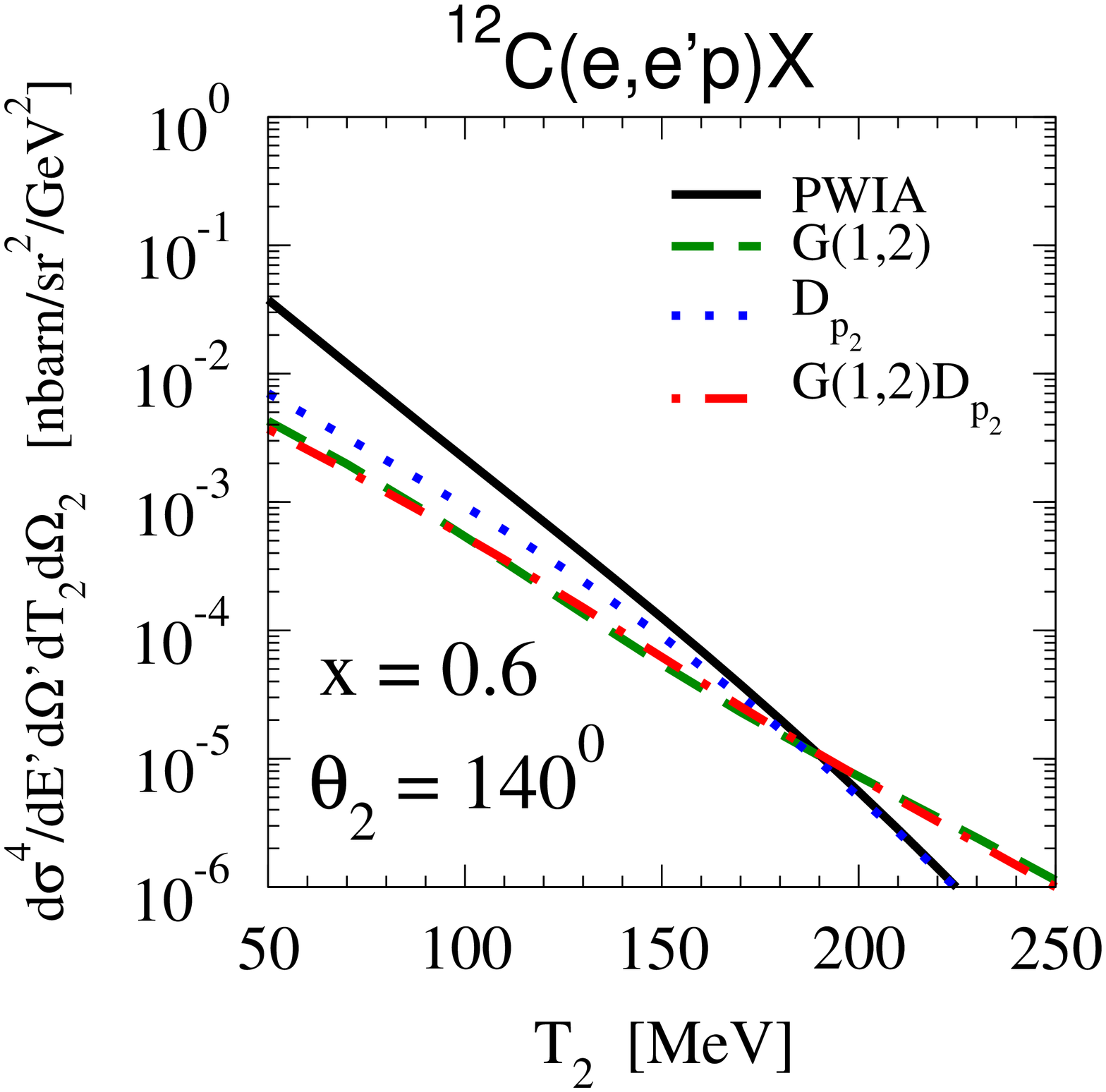}}
\caption{The SIDIS differential cross section for the process $^{12}C(e,e'p)X$ versus
the kinetic energy $T_2$ of the detected proton, emitted backward at $\theta_2=140^0$
for two values of the Bjorken variable. The labeling  of the
different curves is as in Fig.~\ref{contrsep}.}
\label{contrsep2}
\end{figure}

\begin{figure}[!hc]         
\centerline{\hspace{2cm}
\includegraphics[scale=0.33,angle=-90]{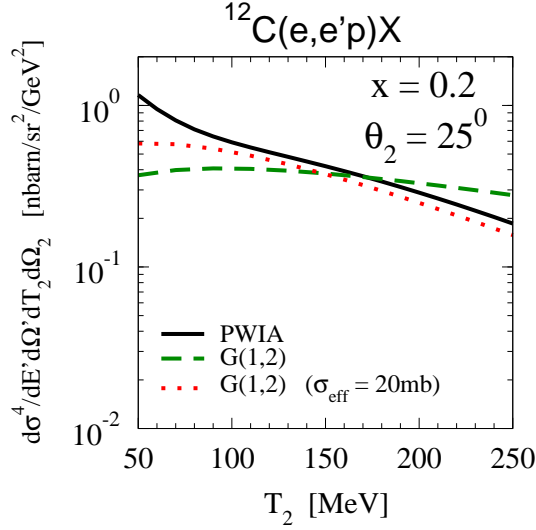}}
\caption{The SIDIS differential cross section for the process $^{12}C(e,e'p)X$ versus the
kinetic energy $T_2$ of the detected proton, emitted forward at $\theta_2=25^0$ and $x=0.2$.
The FSI between the quark-gluon debris and the spectator nucleon is calculated with the
time dependent $\sigma_{eff}(z-z')$ \cite{ciokop} (dashed curve) and with
a costant $\sigma_{eff}=20mb$ (dotted curve). The PWIA is given by the
full curve.}
\label{20mb}
\end{figure}

\begin{figure}[!ht]                
\centerline{\hspace{2cm}
\includegraphics[scale=0.33,angle=-90]{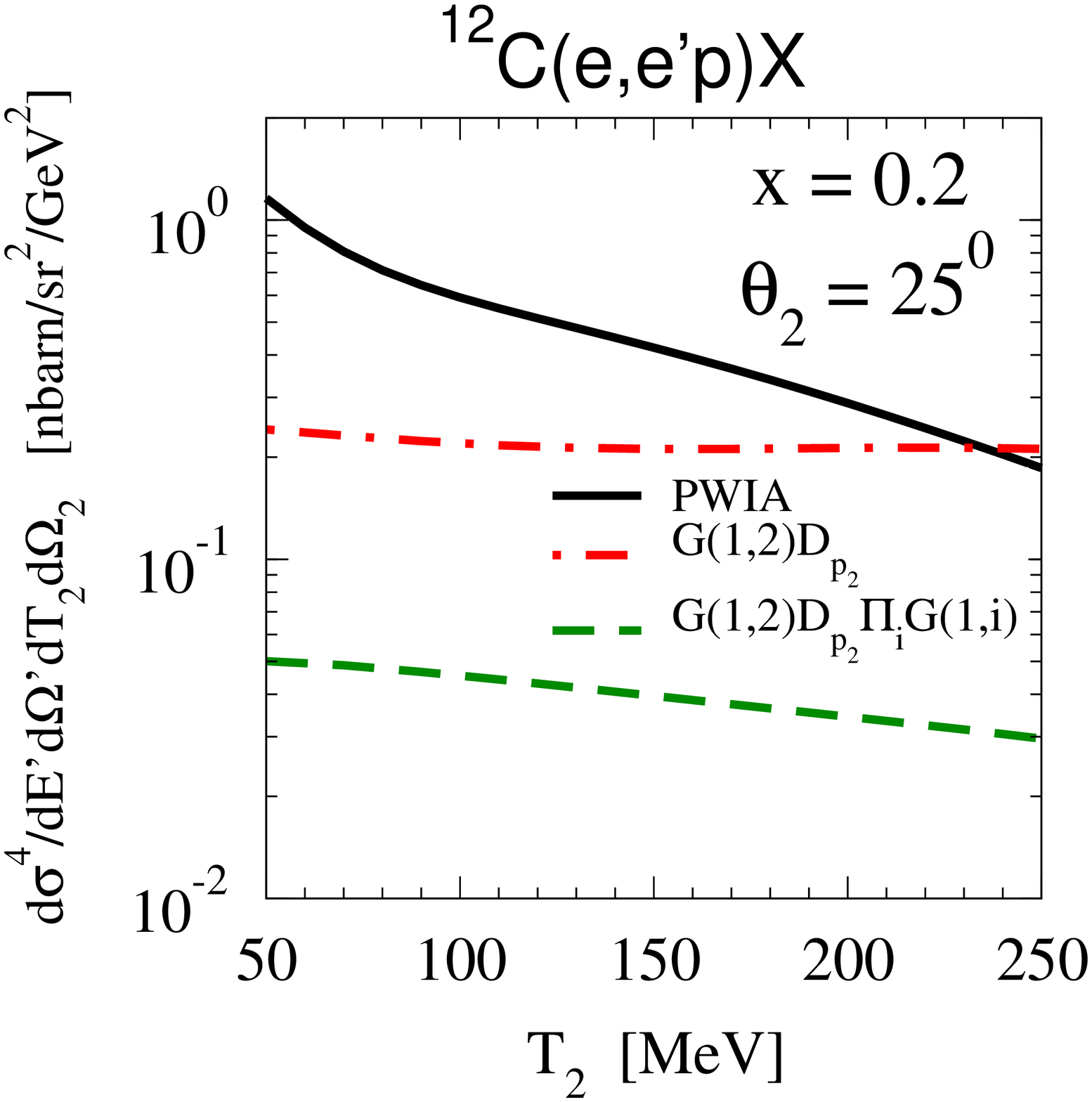}
\hspace{-2cm}
\includegraphics[scale=0.33,angle=-90]{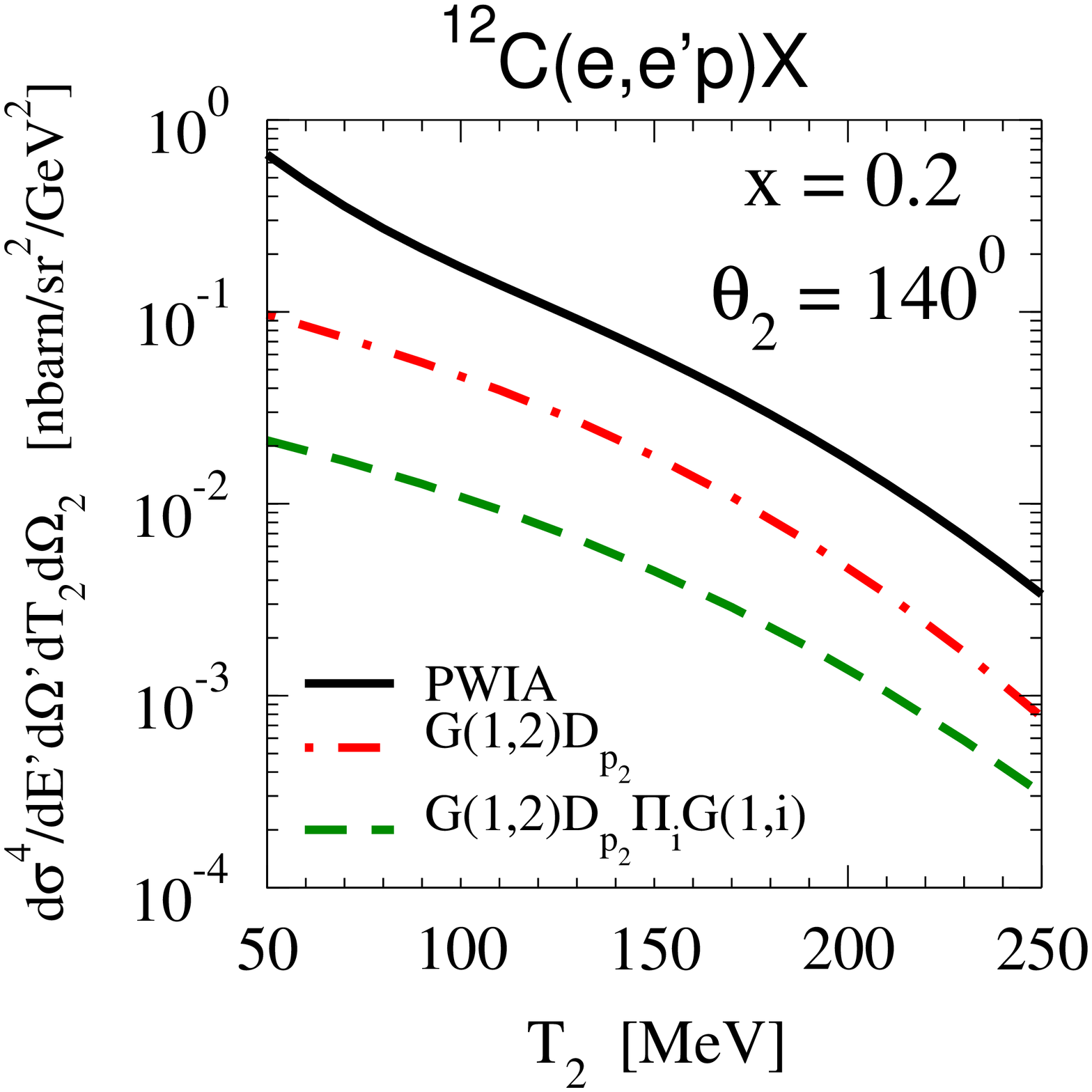}}
\caption{The SIDIS differential cross section for the process $^{12}C(e,e'p)X$ versus the
 kinetic
energy $T_2$ of the detected proton, emitted forward ($\theta_2=25^0$) and backward
($\theta_2=140^0$) at $x=0.2$. The dot-dashed curve includes the FSI between the
quark-gluon debris and the spectator nucleon and between the latter and $A-2$. The
dashed curve includes also the FSI between quark-gluon debris and $A-2$.
The PWIA is given by the solid curves.}
\label{fsitot}
\end{figure}

\begin{figure}[!ht]
\includegraphics[scale=0.33,angle=-90]{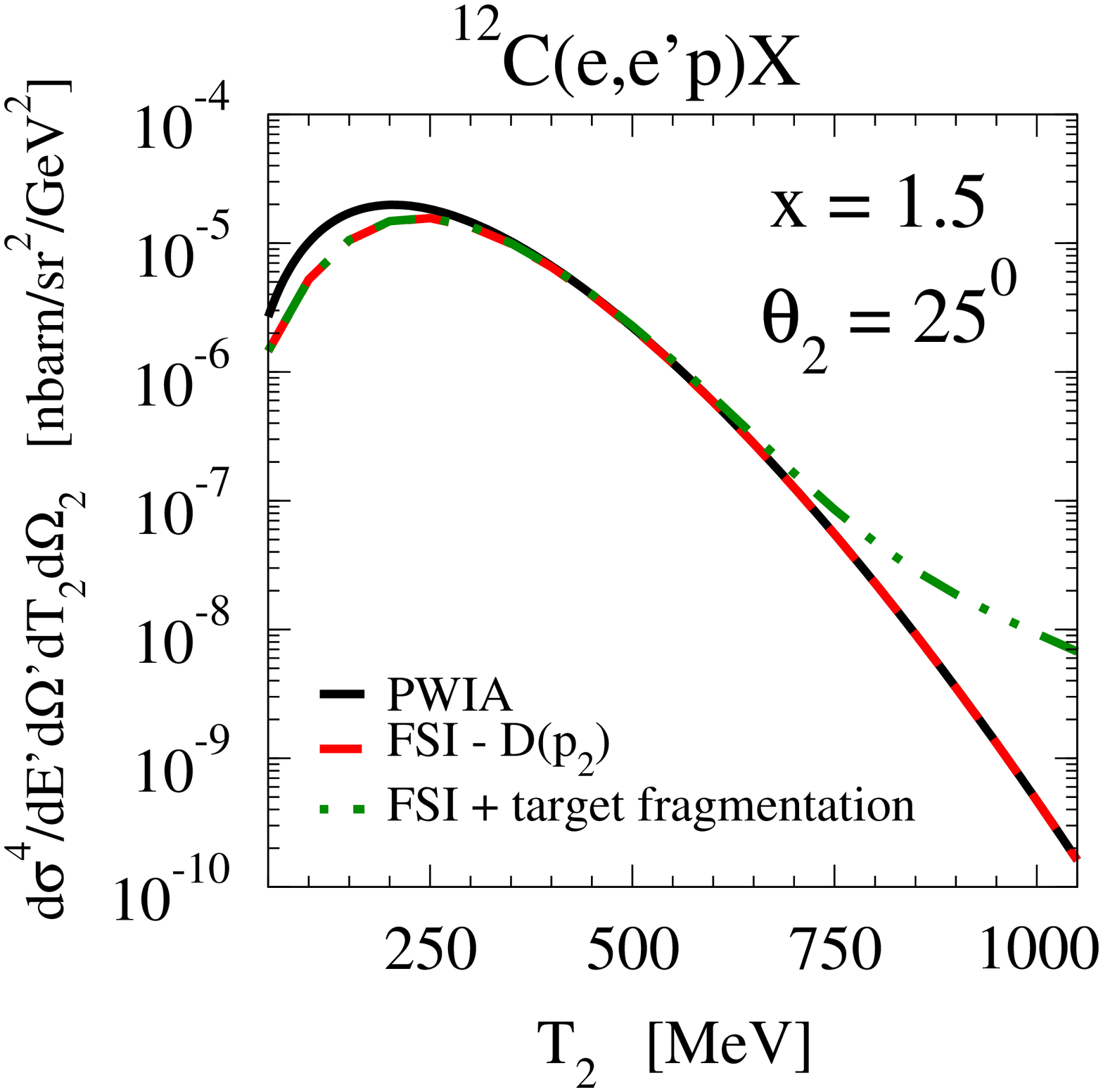}
\caption{The SIDIS differential cross section for the process $^{12}C(e,e'p)X$ versus
the kinetic
energy $T_2$ of the detected proton, emitted forward at $\theta_2=25^0$ and $x=1.5$.
Solid curve: spectator mechanism within the PWIA;
dashed curve: spectator mechanism within the PWIA plus FSI of the
spectator nucleon with $A-2$;
dot-dashed curve: spectator mechanism within the PWIA plus FSI of the
spectator nucleon with $A-2$ and target fragmentation.}

\label{fragcarb}
\end{figure}

We have calculated the differential cross section $d\sigma^A/d\CE'd\Omega'dT_2d\Omega_2$ for
the $^{12}C(e,e^\prime p)X$ process with full FSI described by the operator $\hat{S}_{FSI}$
of
Eq. (\ref{fsiop}).
The results of our calculations are presented in Figs. \ref{contrsep}-\ref{fsitot}, where
the separate  contributions of the various FSI's are shown as a function of the detected
proton kinetic energy $T_2$.
Calculations have been performed assuming an incident electron energy of
$\CE=20\,GeV$ and an electron scattering angle of $\theta_e=15^0$, with values of the
Bjorken scaling variable equal to $x=0.2$ and $0.6$; the proton emission angle has been
fixed at the two values of $\theta_2 = 25^o$ (\textit{forward}) and $\theta_2 = 140^o$
(\textit{backward}).
It can be seen that the most relevant contribution to FSI comes from
hadronization of the hit quark, in forward as well as in backward nucleon emission.
The effects of FSI between the recoiling nucleon and $A-2$
amounts to an attenuation factor which, in the analyzed proton momentum $|\bp_2|$ range,
decreases the cross section by a factor of $\sim$ $0.5-0.7$; as expected, this
contribution is
more relevant for low values of the proton kinetic energy.
We checked the sensitivity of the process upon the nucleon debris effective
cross section using a constant cross section $\sigma_{eff} = 20$ $mb$ \cite{review};
the results presented in Fig. \ref{20mb}  show an appreciable difference with respect to
the calculation performed using the time-dependent effective cross section of ref.
\cite{koppred}.
The results for the differential cross section with the full FSI are presented in Fig.
\ref{fsitot}, for the detection of the slow proton both in the forward
($\theta_2 \,=\, 25^0\;,\; z_2\,<\,1$) and backward ($\theta_2 \,=\, 140^0\;,\; z_2\,>\,1$)
hemispheres.
We analyzed the role of the fragmentation mechanism in the considered SIDIS processes.
The results for the  $^{12}C$ target  are presented
in Fig. \ref{fragcarb}:
it can be seen that, as in the deuteron case, the target fragmentation
mechanism contributes to nucleon emission in the  forward  direction
and becomes noticeable  at high values of $T_2$
($T_2>600$ $MeV$). It should be noted  that such large kinetic
energy are beyond of applicability of our approach. In the region $50$ $MeV<T_2<250$ $MeV$,
where the use of a non relativistic spectral function is well grounded, the effects
of target
fragmentation
play a minor role and the contribution
from final state interaction of the spectator nucleon with $A-2$
practically reduce to an attenuation factor.

\section{Summary and Conclusions}
\n
 We have considered  proton production in   semi inclusive
 processes $A(e, e' p)X$ in the deep inelastic
 limit within the spectator and the target fragmentation mechanisms taking all kind of FSI into account.
 A systematic study of this  process is of  great relevance in hadronic physics.
  As a matter of fact,
 in case of  deuteron targets detailed information on the DIS neutron structure function could in
  principle
 be obtained by
 performing experiments in the kinematical region where FSI are minimized
 (backward production);
  at the same time
 if the experiment is performed when FSI are  maximized (perpendicular kinematics)
  the non   perturbative QCD phenomenon of
  hadronization could be investigated.
 Experimental study of SIDIS off deuteron targets has already
 started at Jlab and is planned to be extended to higher energies at the upgraded facility.

In case of complex nuclei SIDIS  could  also represent a tool  to investigate
 short-range correlations in nuclei.
 As a matter of fact
 the main sources of slow backward protons in SIDIS originates from a correlated pair.

 In the present paper we have investigated in details the effects of FSI.
 In particular, the interaction of the quark-gluon debris with the recoiling proton and the residual
 $A-2$ nucleus, treated within the eikonal approximation,
 and  the interaction of the proton
 with the $A-2$ treated within the distorted wave impulse approximation.
 We have demonstrated, that the FSI due to the quark-gluon debris  off the spectator $A-2$ nucleons
 appreciably attenuates the cross section, reflecting the fact that the rescattering of the quark-gluon
 debris in the final state strongly decreases the
 survival probability of the $(A-2)$ nucleus \cite{ciokop1}.

To main results obtained in the present paper can be summarized as follows:

\begin{enumerate}
\item
in case of SIDIS off the deuteron,  FSI can be  minimized in the parallel
kinematics and maximized in the perpendicular one. In the  former case
the bound nucleon structure function can  be investigated, whereas in the second
case  information on QCD hadronization mechanisms could be obtained;
\item
for complex nuclei,
FSI appreciably decreases the cross section casting doubts as to the possibility
to perform experiments of the type we have considered, where the underlying mechanism is almost
fully exclusive, being the unobserved $(A-2)$ nucleus in a well defined energy state.
   A more realistic case would be to consider a really semi-inclusive
  process by summing over all energy states of $(A-1)$. Calculations of this type are in progress and
  will be presented elsewhere \cite{alvmarchpalli};
\item
as in ref. \cite{ciosim2}, we found that the interaction of the recoiling proton with the $A-2$
system
is relevant only at low proton kinetic energies, leading to an overall small reduction of the cross section;
\item
the target fragmentation mechanism
plays a minor role in slow proton production in SIDIS  except at the
extreme forward direction and large values of the proton momentum.
\end{enumerate}

In conclusion, slow hadron production in SIDIS appears to be a powerful tool to investigate
both the properties of bound nucleons and the  hadronization mechanisms.

\acknowledgments{This work is supported in part by
 the Italian Ministry of Research and University.
 L.P.K. is indebted to  the University of
 Perugia and INFN, Sezione di Perugia, for warm hospitality and financial support. Two of us
 (CdA and LPK)
 have benefitted from illuminating discussions with B. Kopeliovich and S. Kuhn}

%
\newpage

\begin{thebibliography}{15}
%
\bibitem{FS}
           L. L. Frankfurt and M. I. Strikman, Phys. Rep. {\bf 76} (1981) 216;
           160 (1988) 235
\bibitem{review}
                        G. D. Bosveld, A. E. L. Dieperink, O. Scholten, Phys. Rev. {\bf C54}
                        (1989) 79;
                        S. Simula, Phys. Lett. {\bf B387} (1996) 245;
\bibitem{ciosim2}
                 C. Ciofi degli Atti and S. Simula,
                  Phys. Lett. {\bf B319} (1993) 23;
                   C. Ciofi degli Atti and S. Simula, Few-Body Systems  {\bf 18} (1995) 55.

\bibitem{experiment}
                    E. Matsinos, et al., Z. Phys. {\bf C44} (1989) 79\\
                    T. Kitagaki, et al., Phys. Lett. {\bf B214} (1988) 281; G. Guy et al,
                    Phys. Lett. {\bf B229} (1989) 421; M. R. Adams et al, Phys. Lett.
                     {\bf B319} (1993) 23.
\bibitem{simula}
               S. Simula, Phys. Lett. {\bf B387} (1996) 245.
\bibitem{sarg}
             W. Melnitchouk, M. Sargsian and  M. I. Strikman, Z. Phys. {\bf A359} (1997) 99;\\
             M.M. Sargsian, J. Arrington, W. Bertozzi, W. Boeglin et al.,
             J. Phys. {\bf G29} (2003) R1.
\bibitem{scopetta}
                C. Ciofi degli Atti, L. P. Kaptari and S. Scopetta,
                Eur. Phys. J. {\bf A5} (1999) 191.
\bibitem{bonus}
                 S.E. Kuhn et al, JLAB, E94-102 CLAS Colab.,
                 http://www.jlab.org/Hall-B/experiments.
\bibitem{ckk}
                   C. Ciofi degli Atti, L. P. Kaptari, B.Z. Kopeliovich,
                   Eur. Phys. J. {\bf A19} (2004) 145.
\bibitem{boffi}
               S. Boffi, C. Giusti and F.D. Pacati, Phys. Rep. {\bf 226} (1993) 1.
                  Phys. Rev \textbf{C 63} (2001) 044601.
\bibitem{ciokop}
                C.Ciofi degli Atti and B.Kopeliovich,
                Eur. Phys.J. {\bf A17} (2003) 133.
\bibitem{Ffunction}
                   L. Trentadue and G. Veneziano,  Phys. Lett. {\bf B323} (1994) 201.
%
\bibitem{distrtr}
                  S. L. Wu, Phys. Rep. \textbf{107} (1984) 59.

\bibitem{barframaj}
                 A. Bartl, H. Fraas, W. Majerotto, Phys. Rev. D \textbf{26} (1982) 1.
\bibitem{noi}
              M. Alvioli, C. Ciofi degli Atti and V. Palli,
              Nucl. Phys. A \textbf{782} (2007) 175c.
\bibitem{ciosim}
              C. Ciofi degli Atti and S. Simula, Phys. Rev. {\bf C53} (1996) 1689.
\bibitem{KuhnPhysRev}
                    A.V. Klimenko, S.E. Kuhn, C. Butuceanu, K.S. Egiyan et al.,
                    Phys. Rev. {\bf C73} (2006) 035212.
\bibitem{nash_Kuhn_paper}
        M. Alvioli, C. Ciofi degli Atti, L.P. Kaptari,  A.V. Klimenko, S.E. Kuhn,
        V. Palli,    {\it to be published}.
\bibitem{glau1}
              R. J. Glauber, Phys. Rev. \textbf{100} (1955) 242.

\bibitem{glau2}
             R. J. Glauber, High-energy Collision Theory, in W. E. Brittin and L. Dunham,
             editor,
              \emph{Lectures in Theoretical physics} ,
              Ed. W. Brittin, N. Y. Interscience 1959;\\
              R. J. Glauber, ``\emph{High Energy Physics and Nuclear Structure}'', Ed. G.
              Alexander, North Holland, 1967; Ed. S. Devons, Plenum Press, 1970.
\bibitem{ciokop1}
                    C.Ciofi degli Atti and B.Kopeliovich,
                    Phys. Lett. {\bf B606} (2005) 281.
\bibitem {arn}
                  R. A. Arndt et al. ``(SAID) partial wave analisys facility'',
                   http://said.phys.vt.edu
\bibitem{koppred}
              B. Z. Kopeliovich, J. Nemchik, E. Predazzi and A. Hayashigaki, Eur. Phys. J. A
              \textbf{19S1} (2004) 111
\bibitem{alvmarchpalli}
                  M. Alvioli, C. Ciofi degli Atti, I. Marchino and V. Palli, {\it In preparation}
\end{thebibliography}
\end{document}